\documentclass[aps,pre,showpacs,amsmath,amsfonts,amssymb,superscriptaddress,preprint]{revtex4-1}

\usepackage{amsmath}
\usepackage{amsfonts}
\usepackage{amssymb}
\usepackage[english]{babel}
\usepackage[pdftex]{graphicx}
\usepackage{tikz}
\usepackage{rotating}
\usepackage{color}
\usepackage[pdftex,hypertexnames=true,linktocpage=true,breaklinks=true]{hyperref} 
\usepackage[hyphenbreaks]{breakurl}
\hypersetup{colorlinks=true,linkcolor=blue,anchorcolor=blue,citecolor=magenta,filecolor=blue,urlcolor=blue,bookmarksnumbered=false,pdfview=FitB}

\begin{document}




\newcommand{\beq}{\begin{equation}}
\newcommand{\eeq}{\end{equation}}
\newcommand{\p}{\partial}
\newcommand{\Hess}{\mathrm{Hess}}
\newcommand{\bi}{\begin{itemize}}
\newcommand{\ei}{\end{itemize}}
\newcommand{\D}{\mathrm{d}}
\newcommand{\bv}{\bar{v}}
\newcommand{\Christoffel}[3]{\Gamma^{#1}_{#2 #3}}
\newcommand{\ChristoffelTilde}[3]{\widetilde{\Gamma}^{#1}_{#2 #3}}
\newcommand{\LevelSetfunc}[2]{\Sigma_{#1}^{#2}}
\newcommand{\Mfunc}[2]{M_{#1}^{#2}}
\newcommand{\vb}{\bar{v}}
\newcommand{\gdep}[2]{#1_{g_{#2}}}

\newcommand{\intSigmaV}[2]{\int_{\LevelSetfunc{#1}{#2}}}
\newcommand{\bra}[1]{\left\langle #1 \right|}
\newcommand{\ket}[1]{\left| #1 \right\rangle}
\newcommand{\braket}[2]{\left\langle #1  | #2  \right\rangle}

\newcommand{\MultOpH}[1]{\widehat{\mathcal{M}}_{#1}}
\newcommand{\DerOpH}[1]{\widehat{D}_{#1}}
\newcommand{\Adj}[1]{\widehat{#1}^{\dag}}

\newcommand{\C}{\textbf C}
\newcommand{\N}{\mathbb N}
\newcommand{\Z}{\mathbb Z}
\newcommand{\R}{\mathbb R}
\newcommand{\A}{\mathbf A}
\newcommand{\K}{\mathbf K}
\newcommand{\I}{\mathbf I}
\newcommand{\SG}{\mathcal G}
\newcommand{\CC}{\mathrm{C}}
\newcommand{\PC}{\mathcal{P}}
\newcommand{\T}{\mathcal{T}_f}
\newcommand{\Tz}{\mathcal{T}_f^0}

\newcommand{\m}{\mathfrak{M}}
\newcommand{\di}{\mathrm{d}}
\newcommand{\End}{\mathrm{End}}

\newcommand{\Mat}{\mathrm{Mat}}
\newcommand{\GL}{\mathrm{GL}}
\newcommand{\PP}{\mathrm{P}}
\newcommand{\zz}{\mathbb{Z}}
\newcommand{\NN}{\mathbb{N}}
\newcommand{\HH}{\mathrm{H}}
\newcommand{\HHS}{\mathrm{H}^\mathrm{S}}
\newcommand{\ZZ}{\mathrm{Z}}
\newcommand{\BB}{\mathrm{B}}

\newcommand{\Mod}{\mathop{\mathrm Mod}}

\newcommand{\al}{\alpha}
\newcommand{\bt}{\beta}
\newcommand{\om}{\Omega}
\newcommand{\mm}{\omega}
\newcommand{\ups}{\upsilon}
\newcommand{\up}{\Upsilon}
\newcommand{\gm}{ k_1}

\newcommand{\B}{\mathcal{B}}
\newcommand{\M}{\mathcal{M}}
\newcommand{\h}{\mathcal{H}}
\newcommand{\Y}{\mathcal{Y}}

\newcommand{\pp}{\mathcal{P}}
\newcommand{\nn}{\mathcal{N}}

\newcommand{\G}{\mathcal{G}}
\newcommand{\Ord}{\mathcal{O}}

\newcommand{\KK}{\mathfrak{K}}

\newcommand{\gr}{\mathfrak{G}}

\newcommand{\ff}{\mathfrak{F}}
\newcommand{\abs}[1]{\lvert#1\rvert}

\newcommand{\aaa}{\mathcal{A}}
\newcommand{\sss}{\mathcal{S}}
\newcommand{\cl}{\mathrm{Cl}}




\title{On the origin of Phase Transitions in the absence of Symmetry-Breaking}

\author{Giulio Pettini}
\email{pettini@fi.unifi.it}
\affiliation{Dipartimento di Fisica Universit\`a di Firenze, and
I.N.F.N., Sezione di Firenze, via G. Sansone 1, I-50019 Sesto Fiorentino, Italy  \smallskip}

\author{Matteo Gori}\email{gori6matteo@gmail.com}
\affiliation{Centre de Physique Th\'eorique, Aix-Marseille University,
Campus de Luminy, Case 907,
13288 Marseille Cedex 09, France}

\author{Roberto Franzosi}
\email{franzosi@ino.it}
\affiliation{ QSTAR and INO-CNR, largo Enrico Fermi 2,
             I-50125 Firenze, Italy}

\author{Cecilia Clementi}\email{cecilia@rice.edu}
\affiliation{Department of Chemistry, Rice University,
6100 Main street, Houston, TX 77005-1892, USA \smallskip}

\author{Marco Pettini}\email{pettini@cpt.univ-mrs.fr}
\affiliation{Centre de Physique Th\'eorique, Aix-Marseille University,
Campus de Luminy, Case 907,
13288 Marseille Cedex 09, France}

\begin{abstract}

In this paper we investigate the Hamiltonian dynamics of a lattice gauge model in
three spatial dimensions. Our model Hamiltonian is defined on the basis of a continuum version of a 
duality transformation of a three dimensional Ising model. The system so obtained undergoes a
thermodynamic phase transition in the absence of a global symmetry-breaking and thus in the absence of an order parameter.
It is found that the first order phase transition undergone by this model fits into a microcanonical version 
of an Ehrenfest-like classification of phase transitions applied to the configurational entropy.
It is discussed why the seemingly divergent behaviour of
the third derivative of configurational entropy can be considered as the "shadow" of some suitable geometrical and topological transition of
the equipotential submanifolds of configuration space. 
\end{abstract}

\date{\today}


\maketitle

\section{Introduction}
One of the main topics in Statistical Mechanics concerns phase transitions phenomena.  
From the theoretical viewpoint, understanding their origin, and the way of classifying them, 
is of central interest.

Usually, phase transitions are associated with a spontaneous symmetry-breaking phenomenon:  
at low temperatures the accessible states of a system can lack some of the global symmetries of the Hamiltonian, 
so that the corresponding phase is the less symmetric one, whereas at higher temperatures the thermal fluctuations 
allow the access to a wider range of energy states having all the symmetries of the Hamiltonian.
In the symmetry-breaking phenomena, the extra variable which characterizes the physical states of a system is the 
order parameter. The order parameter vanishes in the symmetric phase and is different 
from zero in the broken-symmetry phase. This is the essence of Landau's theory. 
If $G_0$ is the global symmetry group of the Hamiltonian, the order of a phase transition is  determined by the index of the subgroup ${\cal G}\subset G_0$
of the broken symmetry phase. The corresponding mechanism in quantum field theory is described by the Nambu-Goldstone's Theorem.

However, this is not an all-encompassing theory. In fact, many systems do not fit in this scheme and undergo a phase transition in the absence of a symmetry-breaking. This is the case
of liquid-gas transitions,  Kosterlitz-Thouless transitions, coulombian/confined regime transition for gauge theories on lattice,  transitions in glasses and supercooled liquids, 
in general, transitions in amorphous and disordered systems, folding transitions in homopolymers and proteins, to quote remarkable examples. All these physical systems lack
an order parameter.

Moreover, classical theories, as those of Yang-Lee \cite{YL} and of Dobrushin-Lanford-Ruelle \cite{DLR}, require the $N\to\infty$ limit (thermodynamic limit) to 
mathematically describe a phase transition, but the study of transitional phenomena in finite $N$ systems is particularly relevant in many other 
contemporary problems \cite{gross}, for instance related with polymers thermodynamics and biophysics \cite{bachmann}, with Bose-Einstein condensation, 
Dicke's superradiance in microlasers, nuclear physics  \cite{chomaz}, superconductive transitions in small metallic objects. The topological theory of phase transitions provides a natural framework to get rid of the thermodynamic limit dogma because clear topological signatures of phase transitions are found already at finite and small $N$ \cite{book,exact3}.

Therefore, looking for generalisations of the existing theories is a well motivated and timely purpose. The present paper aims at giving a contribution in this
direction along a line of thought initiated several years ago with the investigation of the Hamiltonian dynamical counterpart of phase transitions \cite{book,NCRev,physrep}
which eventually led to formulate a topological hypothesis. 
In fact, Hamiltonian flows (${\cal H}$-flows) can be 
seen as geodesic flows on suitable Riemannian manifolds \cite{book,Pettini}, and then  
the question naturally arises of whether and how these manifolds ``encode'' the fact that their geodesic flows/${\cal H}$-flows 
are associated or not with  a thermodynamic phase transition (TDPT). It is by following this conceptual pathway that one is eventually led to 
hypothesize that suitable topological  changes of certain submanifolds of phase space are the deep origin of TDPT.
This hypothesis was corroborated by several studies on specific exactly solvable models \cite{exact1,exact2,exact3,exact4,santos} and  by 
two theorems. These theorems state that the unbounded growth with $N$ of relevant thermodynamic quantities, eventually leading to singularities in the $N\to\infty$ limit - the hallmark
of an equilibrium phase transition - is {\it necessarily} due to appropriate
topological transitions in configuration space \cite{book,prl1,NPB1,NPB2,kastner}.

Hence, and more precisely, the present paper aims at investigating whether also TDPT occurring in the absence of symmetry-breaking, and thus in the absence of an order parameter, can be
ascribed to some major geometrical/topological change of the previously mentioned manifolds. 

To this purpose, inspired by the dual Ising model, we define a continuous variables
Hamiltonian in three spatial dimensions (3D) having the same local (gauge)
symmetry of the dual Ising model (reported in Section II)
and then proceed to its numerical investigation. The results are reported and discussed in
Section III. Through a standard analysis of thermodynamic observables, it is found that this model undergoes a first order phase transition. It is also found that the larger the number of degrees of freedom the sharper the jump of the second derivative of configurational entropy, what naturally fits into a proposed microcanonical version of an Ehrenfest-like classification of phase transitions.

A crucial finding of the present work consists of the observation that this jump of the second derivative of configurational entropy coincides with a jump of the second derivative of a geometric quantity measuring the total dispersion of the principal curvatures of certain submanifolds (the potential level sets) of configuration space. This is a highly non trivial fact because the peculiar energy variation of the geometry of these submanifolds,  entailing the jump of the second derivative of the total dispersion of their principal curvatures, is a primitive, a fundamental phenomenon: it is the {\it cause and not the effect} of the energy dependence of the entropy and its derivatives, that is, the phase transition is a consequence of a deeper phenomenon. In its turn, the peculiar energy-pattern of this geometric quantity appears to be rooted in the variations of topology of the potential level sets, thus  the present results  provide a further argument in favour of the topological theory of  phase transitions, also in the absence of symmetry-breaking. 

 
\section{The model}
Starting from the Ising Hamiltonian
\begin{equation}
-J\sum_{\langle {\bf i}{\bf j}\rangle \in\Lambda} \sigma_{\bf i} \sigma_{\bf j}
\label{ising}
\end{equation}
with nearest-neighbor interactions ($\langle {\bf i}{\bf j}\rangle$) on a 3D-lattice $\Lambda$, where the $\sigma_{\bf i}$ are
discrete dichotomic variables ($\sigma_{\bf i}= \pm 1$) defined on the lattice sites and $J$ is
real positive (ferromagnetic coupling), one defines the dual
model \cite{kogut}
\begin{equation}
-J \sum_{\square} U_{\bf i j} U_{\bf j k} U_{\bf k l} U_{\bf l i}
\label{dualpotential}
\end{equation}
where the discrete variables $U_{\bf m m^\prime}$ are defined on the links joining the
sites $\bf m$ and $\bf m^\prime$, and $U_{\bf m m^\prime}=\pm 1$.
The summation is carried over all the minimal plaquettes (denoted by $\square$) into which the lattice can be
decomposed.  The dual model in (\ref{dualpotential}) has the local (gauge) symmetry
\begin{equation}
U_{\bf i j} \rightarrow \varepsilon_{\bf i} \varepsilon_{\bf j} U_{\bf i j}
\label{gaugeT}
\end{equation}
with $\varepsilon_{\bf i},\varepsilon_{\bf j}=\pm 1$, and ${\bf i},{\bf j}\in\Lambda$.
Such a gauge transformation leaves
the model (\ref{dualpotential}) unaltered, and after the Elitzur theorem \cite{elitzur}
$\left\langle U_{\bf i j}\right\rangle$ does not qualify as a good order parameter to
detect the occurrence of a phase transition because $\left\langle U_{\bf i j}\right\rangle =0$
always. In other words, no bifurcation of $\left\langle U_{\bf i j}\right\rangle$ can be
observed at any phase transition point inherited by the model (\ref{dualpotential}) from
the Ising model (\ref{ising}).

In order to define a Hamiltonian flow with the same property of local symmetry -- hindering
the existence of a standard order parameter -- we borrow the analytic form of (\ref{dualpotential})
and replace the discrete dichotomic variables $U_{\bf i j}$ with continuous ones
$U_{\bf i j}\in {\mathbb R}$. We remark that we donot want to investigate the dual-Ising model, rather
we just heuristically refer to it in order to define a gauge model with the desired properties.

Moreover, we add to the continuous version of (\ref{dualpotential}) a stabilizing term which is
invariant under the same local gauge transformation (\ref{gaugeT}); this reads
\begin{equation}
\alpha \sum_{\langle{\bf i j}\rangle} \left(U_{\bf i j}^2 - 1\right)^4\ ,
\label{stabilizer}
\end{equation}
where $\langle {\bf i}{\bf j}\rangle$ stands for nearest-neighbor interactions for
link variables and $\alpha$ is a real positive coupling constant.

On a 3D-lattice $\Lambda$, and with $I=\{(1,0,0),(0,1,0),(0,0,1)\}$, we thus define
the following model Hamiltonian
\begin{equation}
{\cal H}(\pi, U) = \sum_{{\bf i}\in\Lambda} \sum_{{\bf \mu}\in I} \frac{1}{2} \pi_{\bf i\mu}^2
-J \sum_{\square\in\Lambda} U_{\bf i j} U_{\bf j k} U_{\bf k l} U_{\bf l i} +
\alpha \sum_{{\bf i}\in\Lambda}\sum_{{\bf\mu}\in I} \left(U_{\bf i\mu}^2 - 1\right)^4
\label{modelH}
\end{equation}
whose flow is investigated through the numerical integration of the corresponding
Hamilton equations of motion.

A more explicit form of (\ref{modelH}) is given by
\begin{eqnarray}
{\cal H}(\pi, U) &= &\sum_{i,j,k=1}^n\sum_{\nu =1}^3 \frac{1}{2}\pi_{ijk\nu}^2 - J\sum_{i,j,k=1}^n
\left[U_{ijk1} U_{i+1 j k 2} U_{i j+1 k 1} U_{i j k 2}\right. \\
&+& \left. U_{ijk2} U_{i j+1 k 3} U_{i j k+1 2} U_{i j k 3}
+ U_{ijk3} U_{i j k+1 1} U_{i+1 j k 3} U_{i j k 1}\right] + \alpha \sum_{i,j,k=1}^n\sum_{\nu =1}^3
\left( U_{i j k \nu}^2 - 1\right)^4\nonumber ,
\label{modelHam}
\end{eqnarray}
where the summation is carried over trihedrals made of three orthogonal plaquettes.
Here $U_{ijk1}$ is the link variable joining the sites $(i,j,k)$ and $(i+1,j,k)$, $U_{ijk2}$ is the
link variable
joining the sites $(i,j,k)$ and $(i,j+1,k)$, $U_{ijk3}$ is the link variable joining the sites
$(i,j,k)$
and $(i,j,k+1)$. Similarly, for example, $U_{i+1jk2}$  joins the sites $(i+1,j,k)$ and
$(i+1,j+1,k)$,  $U_{ij+1k1}$ joins $(i+1,j+1,k)$ and $(i,j+1,k)$, and so on. That is to say
that the fourth index labels the direction, i.e which index is varied by one unit.

The Hamilton equations of motion are given by
\begin{eqnarray}
\dot U_{ijk\nu} & = & \pi_{ijk\nu} , \nonumber \\
\dot \pi_{ijk\nu} & = & {\displaystyle -\frac{\partial {\cal H}}{\partial U_{ijk\nu} }~,~~~~~~~i,j,k=1,\dots,n;~~~ \nu =1,2,3,}
\label{H_dyn_system}
\end{eqnarray}
periodic boundary conditions are always assumed.

The numerical integration of these equations is correctly performed only by means of symplectic
integration schemes. These algorithms satisfy energy conservation (with zero mean
fluctuations around a reference value of the energy) for arbitrarily long
times as well as the
conservation of all the Poincar\'e invariants, which include   
the phase space volume, so that also Liouville's theorem is satisfied by a
symplectic integration. We adopted a third-order bilateral symplectic algorithm as
described in \cite{lapo}. We used $J=1$ and $\alpha = 1$, the integration time step $\Delta t$ varied from $0.005$ at low energy to $0.001$
at high energy so as to keep the relative energy fluctuations $\Delta E/E$ close to $10^{-6}$.

\section{Definition of the observables and Numerical investigation}

Given any observable $A=A(\pi,U)$, one computes its time average as
\beq
\langle A\rangle_t=\frac{1}{t}\int_0^t d\tau\ A[\pi(\tau),U(\tau)]
\label{T-average}
\eeq
along the numerically computed phase space trajectories. For sufficiently
long integration times, and for generic nonlinear (chaotic) systems, these
time averages are used as estimates of microcanonical ensemble averages in
all the expressions given below.

\subsection{Thermodynamic observables}

The basic macroscopic thermodynamic observable is temperature. The
microcanonical definition of temperature depends on entropy --- the basic
thermodynamic potential in the microcanonical ensemble --- according to the relation 
\begin{equation}
\frac{1}{T}=\left(\frac{\partial S}{\partial E}\right)_{\cal V}\ ,
\label{Temperat}
\end{equation}
where $\cal V$ is the volume, $E$ is the energy and the entropy $S$ is given by 

\begin{equation}
S(N,E,{\cal V})=k_{\rm B} \log \int d\pi_1\cdots d\pi_N dU_1\cdots dU_N\ \delta[E - {\cal H}(\pi,U)]
\label{esse1}
\end{equation}
where $N$ is the total number of degrees of freedom, $N=3n^3$ in the present context, and $U_k$ stands for any suitable labelling
of them.
By means of a Laplace transform technique \cite{pearson}, from Eqs. (\ref{Temperat}) and (\ref{esse1})
one gets (setting $k_{\rm B}=1$)
\begin{equation}
T=2 \left[(N-2)\langle K^{-1}\rangle\right]^{-1}~.
\label{Temp2}
\end{equation} 
where $\langle K^{-1}\rangle$ is the microcanonical ensemble average of
the inverse of the kinetic energy $K=E-V(U)$, where $V(U)$ is the potential part of the
Hamiltonian (\ref{modelHam}). 

In numerical simulations 
\begin{equation}
\langle K^{-1}\rangle = \frac{1}{t}\int_0^td\tau\ \left[ \sum_{i,j,k=1}^n\sum_{\nu=1}^3\frac{1}{2}\pi_{ijk\nu}^2(\tau)\right]^{-1}\ .
\label{Temp1bis}
\end{equation}
For $t$ sufficiently large  $\langle K^{-1}\rangle$ attains a stable value (in general
this is a rapidly converging quantity entailing a rapid convergence of  $T$).

Since the invariant measure for nonintegrable Hamiltonian dynamics is the microcanonical
measure in phase space, the occurrence of equilibrium phase transitions can be investigated in the
microcanonical statistical ensemble through Hamiltonian dynamics \cite{book,ergodic}.
Standard numerical signatures of phase transitions (also found with canonical Monte Carlo random
walks in phase space) are: the bifurcation of the order parameter at the
transition point (somewhat smoothed at finite number of degrees of freedom),
and sharp peaks -- of increasing height with an increasing number of degrees
of freedom -- of the specific heat at the transition point.
As already remarked above, our model (\ref{modelHam}), because of the local (gauge) symmetry, lacks
a standard definition of an order parameter as is usually given in the case of
symmetry-breaking phase transitions. And in fact, in every numerical simulation of the dynamics we have computed the time average $\left\langle \left\langle U_{\bf i j}\right\rangle\right\rangle_t$ 
always finding $\left\langle \left\langle U_{\bf i j}\right\rangle\right\rangle_t \simeq 0$ independently of the lattice size and of the energy value (the double average means averaging over the entire lattice first, and then averaging over time). 

Thus, the presence of a phase transition is detected through the shape of the so-called caloric curve, that is, $T = T(E)$. For the model in (\ref{modelHam}) this has been computed 
 by means of Eq. (\ref{Temp2}). Then 
 the {\it microcanonical} constant-volume specific heat follows according to the relation  
$1/C_{\cal V}=\partial T(E)/\partial E$. The numerical computation of specific heat can be independently performed, with respect to the caloric curve, as follows.
Starting with the definition of the entropy, given in (\ref{esse1}), an analytic 
formula can be worked out \cite{pearson}, which is exact at {\it any}
value of $N$.
This formula reads
\begin{equation}
c_{\cal V}(E)=\frac{C_{\cal V}}{N}=
\frac{N(N-2)}{4}\left[(N -2) - (N - 4)
\frac{\langle K^{-2}\rangle}{\langle K^{-1}\rangle^2}\right]^{-1}\ ,
\label{cv2}
\end{equation}
and this is the natural expression to work out the microcanonical specific
heat by means of Hamiltonian dynamical simulations.

In order to get the above defined specific heat,  time averages of the kind
\[
\langle K^\alpha\rangle_t =\frac{1}{t}\int_0^t d\tau\ \left[ \sum_{i,j,k=1}^n\sum_{\nu=1}^3\frac{1}{2}\pi_{ijk\nu}^2(\tau)\right]^\alpha
\]
are computed with $\alpha = -1, -2$. Then, for sufficiently large $t$, the microcanonical averages $\langle K^\alpha\rangle$
can be replaced by $\langle K^\alpha\rangle_t$.

\begin{figure}[h!] \centering
\includegraphics[scale=0.3,keepaspectratio=true]{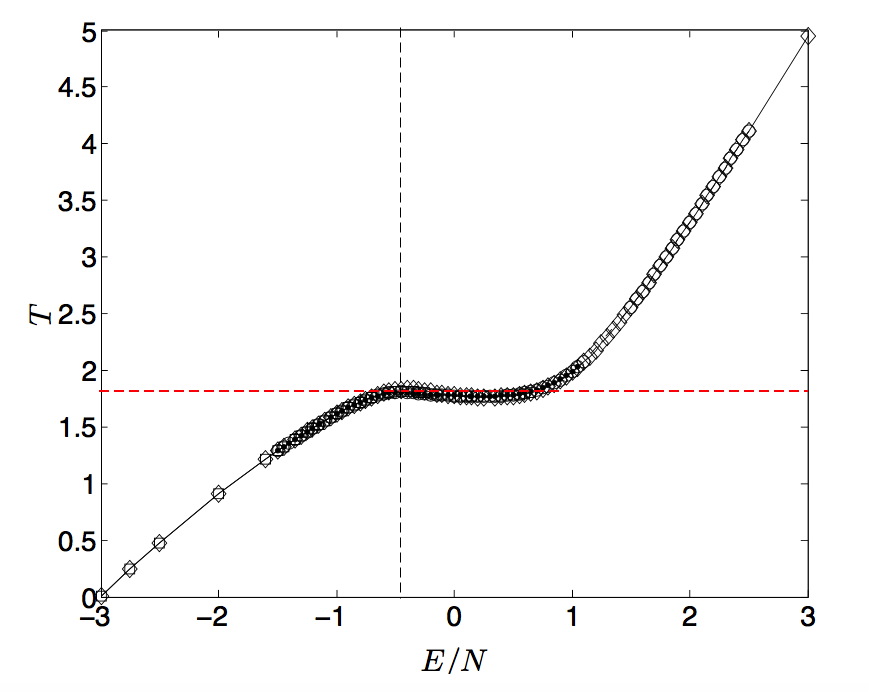} 
\caption[10pt]{(Color online) Caloric curve. Temperature is computed according to Eq.{\protect(\ref{Temp2})}. Lattice dimensions:  $n^3=6\times 6\times 6$ (rhombs),   $n^3=8\times 8\times 8$ (squares), $n^3=10\times 10\times 10$ (circles).
The dashed lines identify the point of flat tangency at lower energy.}
\label{1}
\end{figure}

\begin{figure}[h!]
\includegraphics[width=8cm,angle=-90]{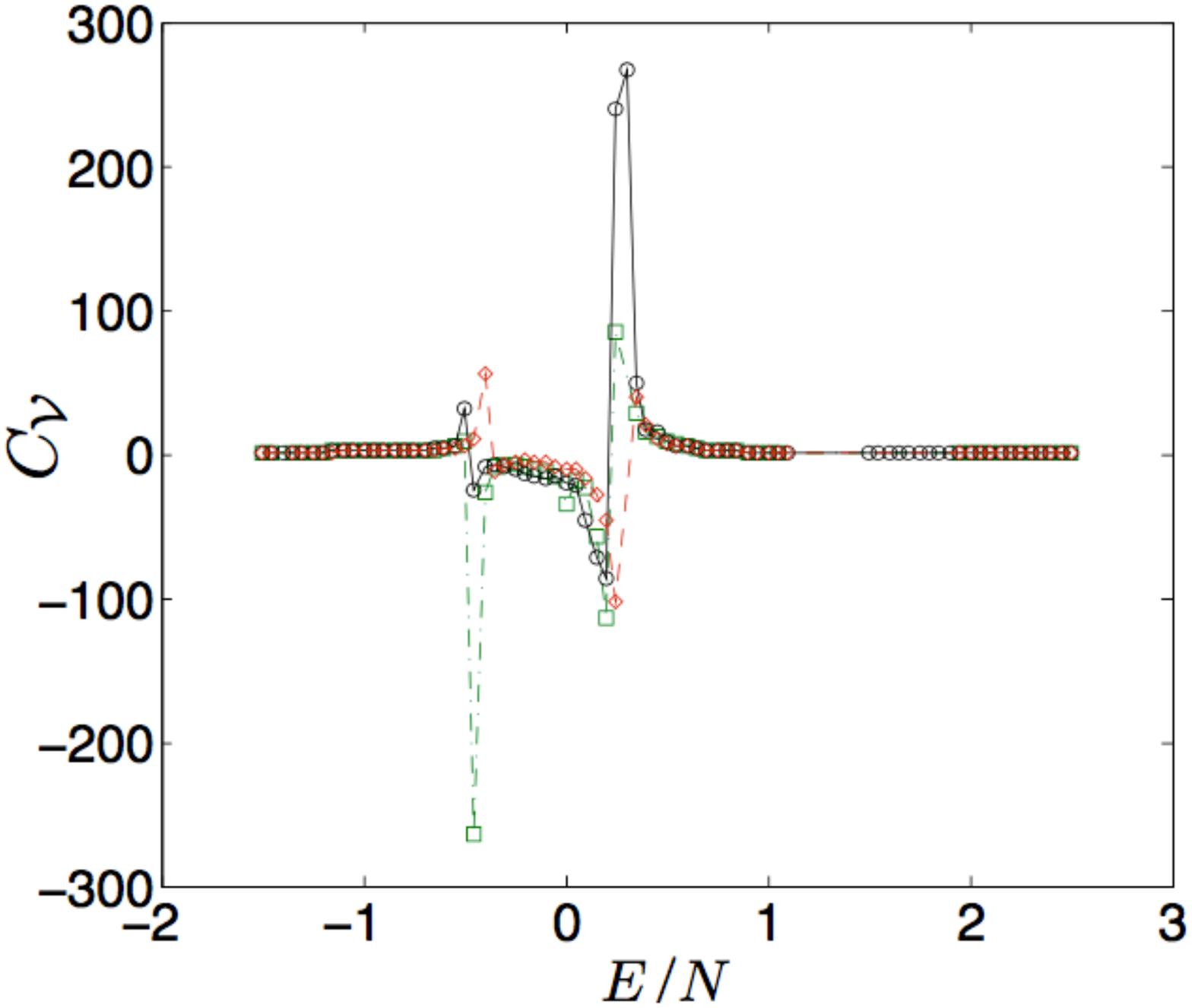}
\caption{(Color online) Constant volume specific heat computed by means of Eq.{\protect{(\ref{cv2})}}.   
Lattice dimensions:  $n^3=6\times 6\times 6$ (rhombs),   $n^3=8\times 8\times 8$  (squares), $n^3=10\times 10\times 10$ (circles). }
\label{cal-spec}
\end{figure}
\begin{figure}[h!]
\includegraphics[width=9cm,angle=0]{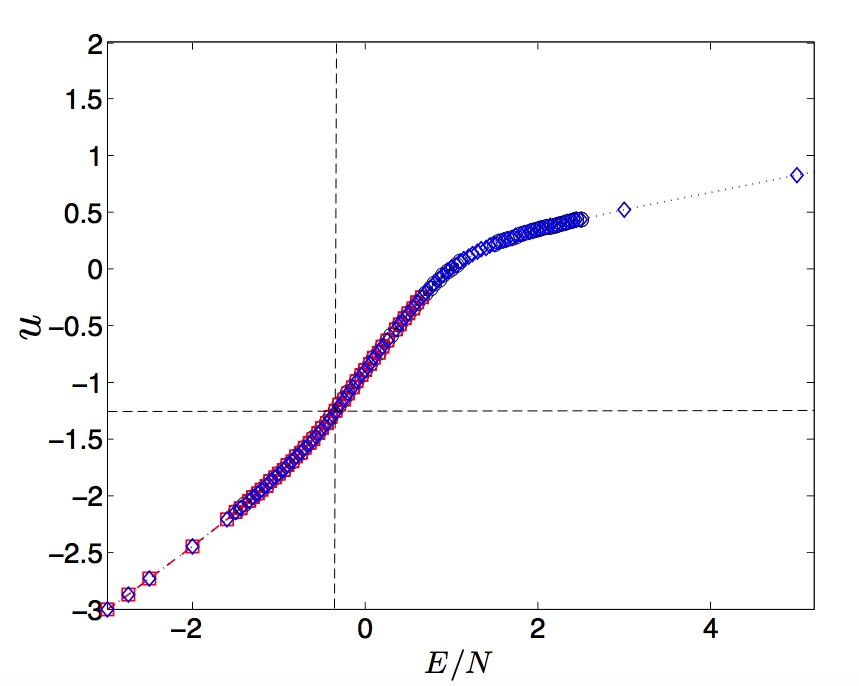}
\caption{(Color online) Internal potential energy density computed through Eq.{\protect{(\ref{T-average})}} where the observable $A$ is the potential function per degree of freedom of the system.
Lattice dimensions:  $n^3=6\times 6\times 6$ (rhombs),   $n^3=8\times 8\times 8$  (squares), $n^3=10\times 10\times 10$ (circles).}
\label{intern-energ}
\end{figure}

In \autoref{1} the caloric curve is reported for different sizes of the lattice. A kink, typical of first order phase transitions, can be seen. This
entails the presence of negative values of the specific heat, and, consequently, ensemble nonequivalence for the model under consideration.
And, in fact, in \autoref{cal-spec}, where we report the outcomes of the computations of the specific heat according to Eq.(\ref{cv2}), we can observe an energy interval where
the specific heat $C_{\cal V}$ is negative, and very high peaks are also found. Nevertheless, these peaks are not related with an analyticity loss of the entropy (see Section \ref{IIIC}) but rather depend on the existence of two points of flat tangency to the caloric curve. 

In \autoref{intern-energ} the average potential energy per lattice site $u = \langle V\rangle/N$ is displayed as a function of the total energy density. Also in this case we observe a regular function which is stable with the number of degrees of freedom. The dashed lines identify the phase transition point which corresponds to $E_c/N\simeq -0.40$ and  $u_c/N\simeq -1.32$.

\color{black} The results so far reported provide us with a standard numerical evidence of the existence of a first order phase transition undergone by the model investigated. \color{black}
Besides standard thermodynamic observables, the study of phase transitions through Hamiltonian dynamics makes available a new observable, 
the largest Lyapunov exponent $\lambda$, which is of purely dynamical kind, and which has usually displayed peculiar patterns in presence 
of a symmetry-breaking phase transition \cite{book,physrep,cccp,cccppg,lando,firpo,dauxois}. Therefore in the following Section an attempt is made to characterise the phase transition undergone by our model also through the energy dependence of $\lambda$. 
 
\subsection{A Dynamic observable: the largest Lyapunov exponent}
The largest Lyapounov exponent $\lambda$ is
the standard and most relevant indicator of the dynamical stability/instability (chaos)
of phase space trajectories. Let us quickly recall that the numerical computation of
$\lambda$ proceeds by integrating the tangent dynamics equations, which, for Hamiltonian flows, read
\begin{eqnarray}
\dot \xi_i & = & \zeta_i~, \nonumber \\
\dot \zeta_{i} & = & {\displaystyle  -\sum_{j=1}^N
\left(\frac{\partial^2 V}{\partial q_1 \partial q_j}\right)_{q(t)} \xi_j }~,\ \ \ \ \ \ \ \  i=1,\dots,N
\label{H_tgdyn_system}
\end{eqnarray}
together with the equations of motion of the Hamiltonian system under investigation.
Then the largest Lyapunov exponent $\lambda$ is defined by
\beq
\lambda = \lim_{t \to \infty} \frac{1}{t} \log
\frac{\left[\xi_1^2(t) +
\cdots + \xi_N^2(t) + \zeta_1^2(t) +
\cdots + \zeta_N^2(t)\right]^{1/2}}
{\left[\xi_1^2(0) +
\cdots + \xi_N^2(0) + \zeta_1^2(0) +
\cdots + \zeta_N^2(0)\right]^{1/2}} ~,
\label{def_lambda_standard}
\eeq
In a numerical computation the discretized
version of (\ref{def_lambda_standard}) is used, with $\boldsymbol{\xi} =(\xi_1,\dots,\xi_{2N})$ and $\xi_{i + N} = \dot \xi_i $
\beq
\lambda = \lim_{m \to \infty} \frac{1}{m} \sum_{i=1}^m
\frac{1}{\Delta t}
\log\frac{\Vert\boldsymbol{\xi}[(i+1)\Delta t ]\Vert}{\Vert\boldsymbol{\xi}(i\Delta t)\Vert}~,
\label{lambda_num}
\eeq
where, after a given number of time steps $\Delta t$, for practical numerical
reasons it is convenient to renormalize  the value
of $\Vert\boldsymbol{\xi}\Vert$ to a fixed one.
The numerical estimate of $\lambda$ is obtained by retaining the time asymptotic value of $\lambda(m\Delta t)$. This is obtained by checking the relaxation pattern of
$\log \lambda(m \Delta t)$ versus $\log (m \Delta t)$.

\begin{figure}[h!]
\includegraphics[width=9.5cm,angle=0]{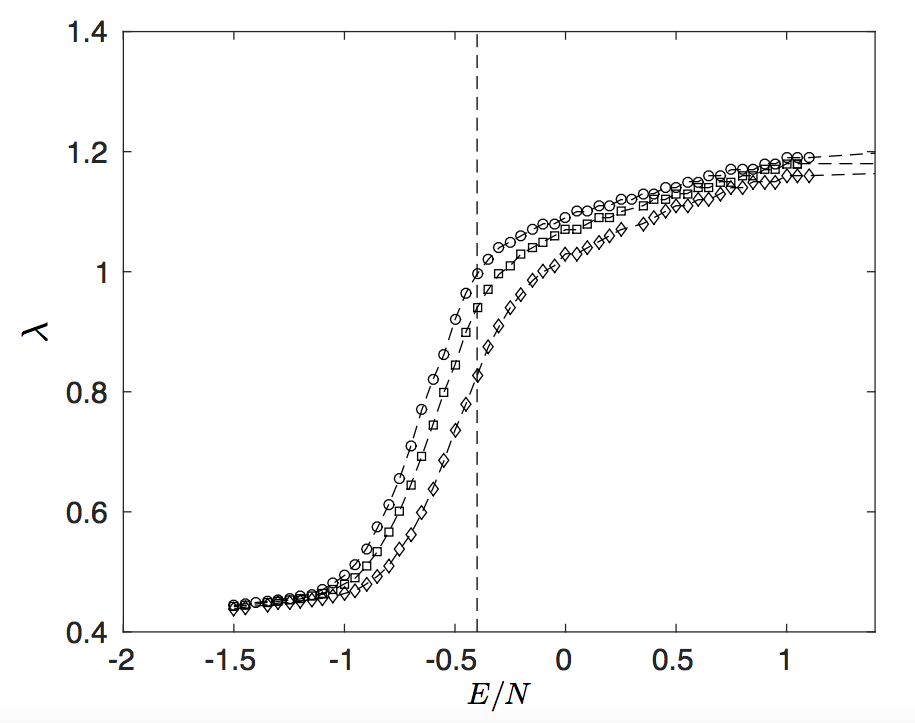}
\caption{Largest Lyapunov exponent versus the energy per degree of freedom.
Lattice dimensions: $n^3=6\times 6\times 6$ (rhombs),   $n^3=8\times 8\times 8$  (squares), $n^3=10\times 10\times 10$ (circles). The dashed vertical	l line indicates the phase transition point.}
\label{lyap-fig}
\end{figure}
Note that $\lambda$ can be expressed as the time average of a suitable observable defined as follows.
From the compact notation
\[
\dot\xi_i = \sum_k J_{ik} [q(t)] \xi_k\, 
\]
for the system (\ref{H_tgdyn_system}) and observing that
\[
\frac{1}{2}\frac{d}{dt}\log (\xi^T\xi)=
\frac{\xi^T\dot\xi + \dot\xi^T\xi}{2\xi^T\xi} =\frac{\xi^T J[q(t)] \xi +
\xi^T J^T[q(t)] \xi }{2\xi^T\xi}\, ,
\]
setting ${\cal J}[q(t),\xi(t)]=\{\xi^T J[q(t)] \xi + \xi^T J^T[q(t)] \xi \}/(2\xi^T\xi)$,
one gets
\begin{equation}
\lambda = \lim_{t\rightarrow\infty}\frac{1}{t} \log \frac{\Vert\xi(t)\Vert}
{\Vert\xi(0)\Vert} = \lim_{t\rightarrow\infty}\frac{1}{t} \int_0^t d\tau\,
 {\cal J}[q(\tau),\xi(\tau)] ~,
\label{lambda_def_av}
\end{equation}
which {\it formally} gives $\lambda$ as a time average as per Eq.(\ref{T-average}).

The numerical results summarised in \autoref{lyap-fig} qualitatively indicate a transition between two dynamical regimes of chaoticity: from a weakly chaotic dynamics at low energy density values, to a more chaotic dynamics at large energy density values. However the transition between these dynamical states is a mild one. At variance with those models where a phase transition stems from a symmetry-breaking, here there is no peculiar property of the shapes of $\lambda = \lambda(E/N)$  in correspondence of the phase transition.
Therefore, in the following Section we directly tackle the numerical study of the differentiability class of the entropy.

\subsection{Microcanonical definition of phase transitions}\label{IIIC}
As is well known, according to the Ehrenfest classification,  the order of a phase  transition is given by the order of the discontinuous derivative with respect to temperature $T$ of the Helmholtz free energy $F(T)$. However, a difficulty arises in presence of divergent specific heat $C_V$ associated with a second order phase transition because this implies a divergence of  $(\partial^2F/\partial T^2)$, and, in turn, a discontinuity of $(\partial F/\partial T)$ so that the distinction between first and second order transitions is lost.
By resorting to the concept of symmetry-breaking, Landau theory circumvents this difficulty by classifying the order of a phase transition according to the index of the symmetry group of the broken-symmetry phase which is a subgroup of the group of the more-symmetric phase. As in the present work we are tackling a system undergoing a phase transition in the absence of symmetry-breaking, we have to get back to the origins as follows. According to the Ehrenfest theory, a phase transition is associated with a loss of analyticity of a thermodynamic potential (Helmholtz free energy, or, equivalently Gibbs free energy), and the order of the transition depends on the differentiability class of this thermodynamic potential. Later, on a mathematically rigorous ground, the identification of a phase transition with an analyticity loss of a thermodynamic potential (in the gran-canonical ensemble) was confirmed by the Yang-Lee theorem. Now, let us consider the statistical ensemble which is the natural counterpart of microscopic Hamiltonian dynamics, that is, microcanonical ensemble. As already recalled in Section III.A, here the relevant thermodynamic potential is entropy, and considering the specific heat
\begin{equation}
C_V^{-1} = \frac{\partial T(E)}{\partial E}\ \qquad {\rm which,\,  after\, Eq.\protect{\eqref{Temperat}}},\, {\rm reads} \qquad C_V = - \left(\frac{\partial S}{\partial E}\right)^{2} \left(\frac{\partial^2S}{\partial E^2}\right)^{-1}\ ,
\end{equation}
from the last expression we see that $C_V$ can diverge only as a consequence of the vanishing of $({\partial^2 S}/{\partial E^2})$ which has nothing to do with a loss of analyticity of $S(E)$. This is why in Section III.A we have affirmed that the peaks of $C_V$ reported in \autoref{cal-spec} stem from a rather trivial effect.
For standard Hamiltonian systems (i.e. quadratic in the momenta) the relevant information is carried by the configurational microcanonical ensemble, where the configurational canonical free energy is 
      \[
    f_N(\beta)\equiv f_N(\beta; V_N)=
        \frac{1}{N} \log Z_c(\beta, N)\,  
    \]
    with 
    \begin{eqnarray}
Z_c(\beta, N) &=& \int_{(\Lambda^d)^{\times n}}dq_1\dots dq_N\ \exp
[-\beta V_N(q_1,\dots, q_N)]\nonumber
\label{Zconf}
\end{eqnarray}
and the configurational microcanonical entropy  (in units s.t. $k_B=1$) is 
\begin{eqnarray}
      S_N(\vb) \equiv S_N(\vb;V_N)
          =\frac{1}{N} \log{ \Omega (N \vb, N)} \ ,\nonumber
\end{eqnarray}
where $\vb = v/N$ is the potential energy per degree of freedom, and
\begin{equation}
\Omega(v,N) = \int_{(\Lambda^d)^{\times n}} dq_1\cdots dq_N\ \delta
[V_N(q_1,\dots, q_N) - v] ~, \label{pallaM}
\end{equation}
 is the volume of the equipotential hypersurface $\Sigma_v$ (codimension-$1$ subset of configuration space), and $\delta[\cdot]$ is the Dirac functional.
Then $S_N(\vb)$ is
related to the configurational canonical free energy, $f_N$, for
any $N\in{\mathbb N}$, $\vb\in{\mathbb R}$, and $\beta \in{\mathbb R}$ through the Legendre
transform 
    \begin{eqnarray}
     - f_N(\beta) =  \beta \cdot \vb_N -  S_N(\vb_N)
    \label{legendre-tras}
    \end{eqnarray}
where the inverse of the configurational temperature $T(v)$  is given by
      \begin{eqnarray}
    \beta_N(\vb)=
    \frac{\partial S_N}{\partial \vb} (\vb) \, .
    \label{vb2beta}
    \end{eqnarray}
Then consider the function $\phi(\vb)=f_N[\beta(\vb)]$, from $\phi^\prime(\vb) = -\vb\  [d \beta_N(\vb)/d\vb]$
we see that if $\beta_N(\vb)\in{\cal C}^k(\mathbb R)$ then also $\phi(\vb)\in{\cal C}^k(\mathbb R)$ which in turn
means $S_N(\vb)\in{\cal C}^{k+1}(\mathbb R)$ while $f_N(\beta)\in{\cal C}^k(\mathbb R)$. 

Hence, if the functions $\{S_N(\vb)\}_{N\in{\mathbb N}}$ are convex, thus ensuring the existence of the above Legendre transform, and if in the $N\to\infty$ limit it is $f_\infty(\beta)\in{\cal C}^0(\mathbb R)$ then $S_\infty(\vb)\in{\cal C}^1(\mathbb R)$, and if 
$f_\infty(\beta)\in{\cal C}^1(\mathbb R)$ then $S_\infty(\vb)\in{\cal C}^2(\mathbb R)$. We are now ready for a classification of phase 
transitions {\it \`a} {\it la} Ehrenfest in the present microcanonical configurational context.

The original Ehrenfest definition associates a first or
second order phase transition with a discontinuity in the first or second derivatives of $f_\infty(\beta)$, that is with $f_\infty(\beta)\in{\cal C}^0(\mathbb R)$ or 
$f_\infty(\beta)\in{\cal C}^1(\mathbb R)$, respectively. This premise heuristically suggests to associate a first order phase transition with a discontinuity of the second derivative of the 
entropy $ S_\infty(\vb)$, and to associate a second order phase transition with a discontinuity of the third derivative of the entropy $S_\infty(\vb)$. 
Let us stress that this definition is proposed regardless the existence of the Legendre transform, which typically fails in presence of first order phase transitions which bring about a kink-shaped energy dependence of the entropy \cite{gross}. Thus, strictly speaking, the definition that we are putting forward does not mathematically and logically stem from the original Ehrenfest classification.  The introduction of our entropy-based classification of phase transitions  {\it \`a} {\it la} Ehrenfest is heuristically motivated, but to some extent arbitrary. This entropy-based classification no longer suffers the previously mentioned difficulty arising in the framework of canonical ensemble, including here both divergent specific heat in presence of a second order phase transition and ensemble non-equivalence. In the end the validity of the proposed classification has to be checked against practical examples.
The gauge model, here under investigation, provides a first benchmarking  in this direction. 

{\color{black} It is worth mentioning that a thorough investigation of  microcanonical  thermodynamics, with special emphasis on phase transitions, can be found in Ref.\cite{gross}. However, our above proposed approach is rather different from (and complementary to) that discussed in \cite{gross}. In fact, in \cite{gross} the determinant and the eigenvalues of the curvature matrix associated with the two dimensional entropy surface $S(E,N)$ are the basic quantities to signal the presence, and define the order, of a phase transition; while no reference is made to the differentiability class of the entropy to characterise a phase transition.          }

\begin{figure}[h!]
\includegraphics[width=10cm,angle=0]{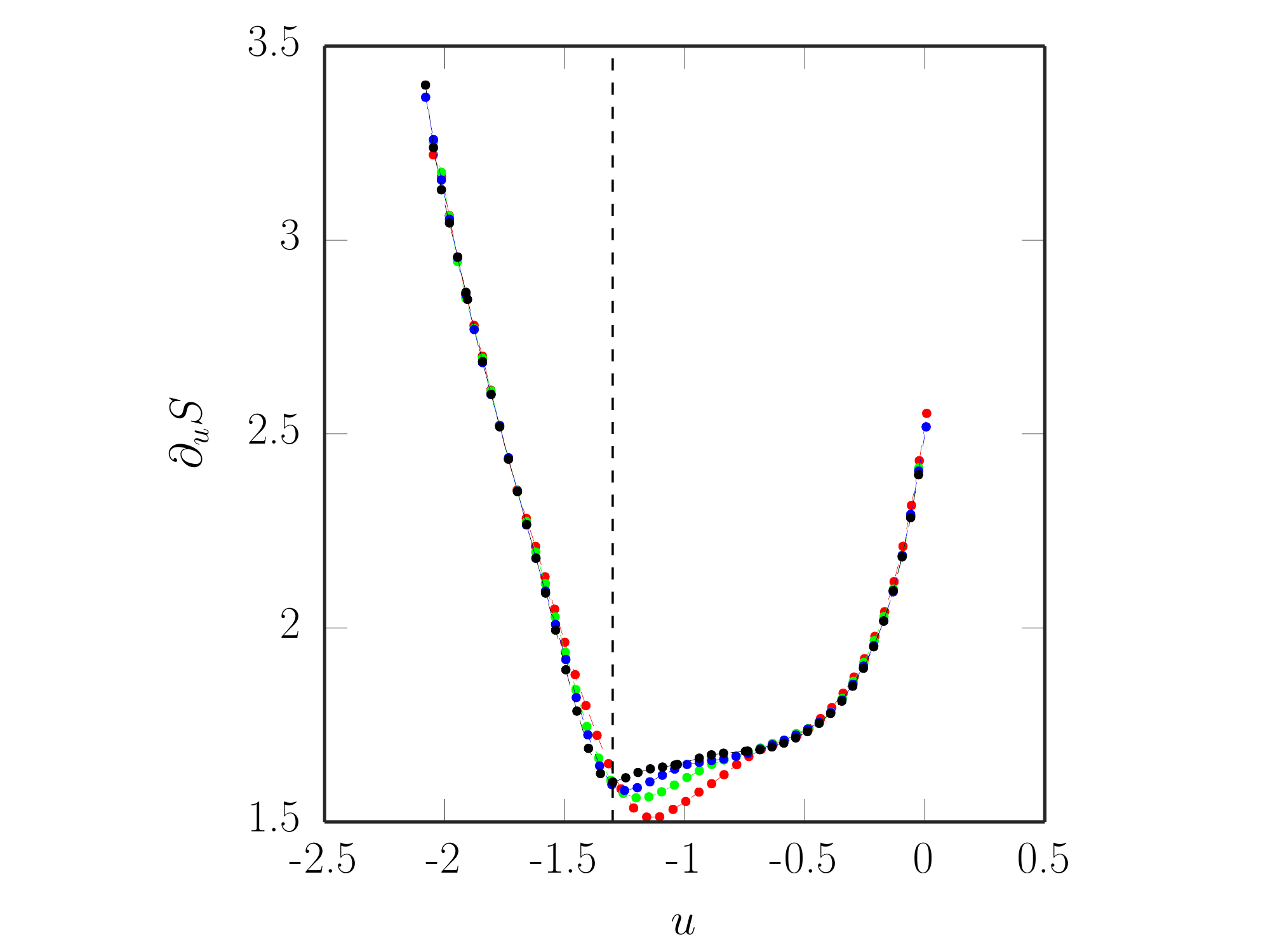}
\caption{(Color online) First derivative $\partial S/\partial u$ of the configurational entropy versus the average potential energy per degree of freedom $u$.
Lattice dimensions: $n^3=6\times 6\times 6$ (red full circles), $n^3=8\times 8\times 8$  (green full circles),   $n^3=10\times 10\times 10$ (blue full circles), $n^3=14\times 14\times 14$ (black full circles). The vertical dashed line locates the phase transition point.}
\label{deriv1S}
\end{figure}

\begin{figure}[h!]
\includegraphics[width=10cm,angle=0]{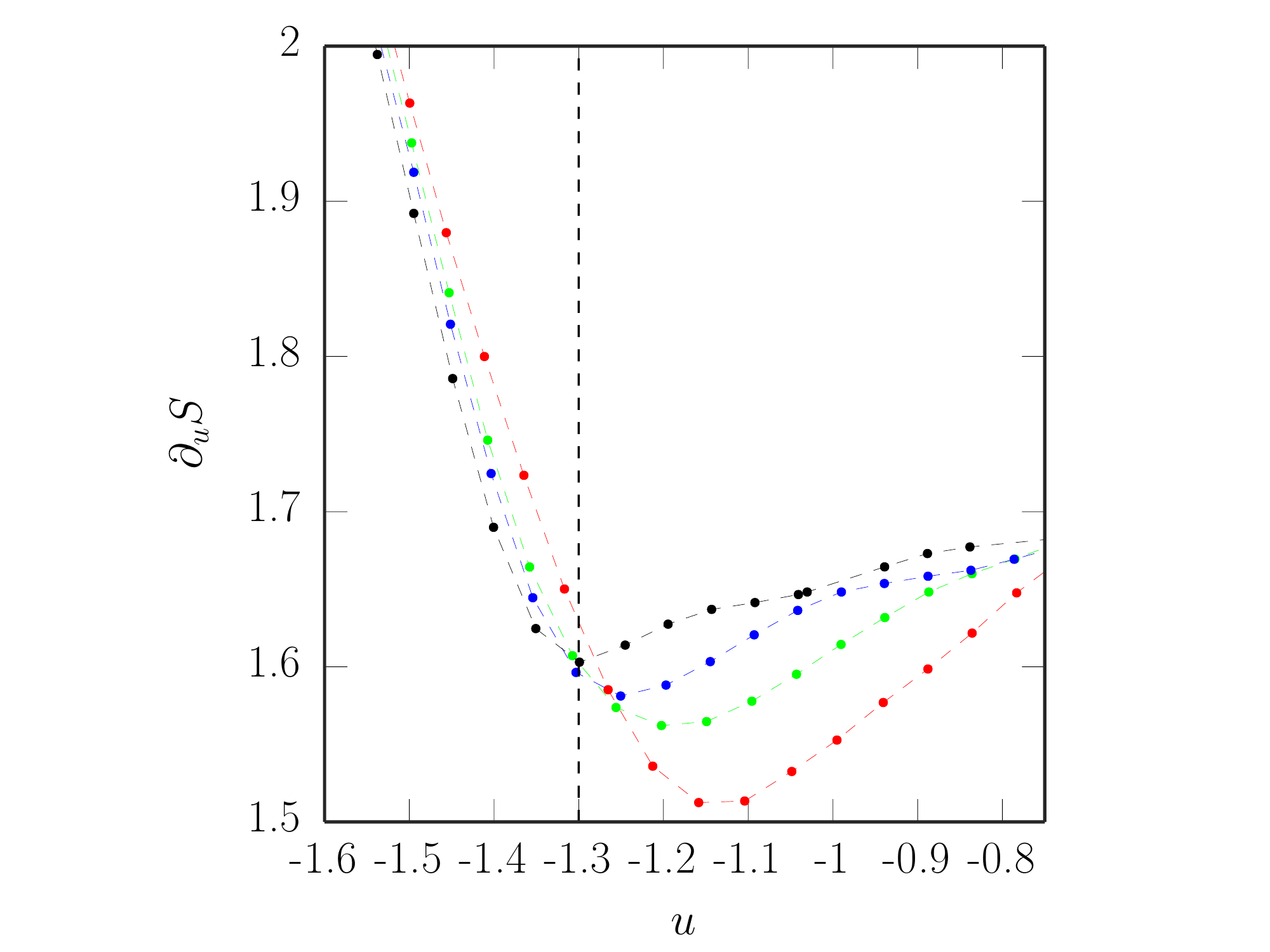}
\caption{(Color online) Zoom of the first derivative $\partial S/\partial u$ of the configurational entropy versus the average potential energy per degree of freedom $u$.
Lattice dimensions: $n^3=6\times 6\times 6$ (red full circles), $n^3=8\times 8\times 8$  (green full circles),   $n^3=10\times 10\times 10$ (blue full circles), $n^3=14\times 14\times 14$ (black full circles). The vertical dashed line locates the phase transition point.}
\label{deriv1Szoom}
\end{figure}

\begin{figure}[h!]
\includegraphics[width=10cm,angle=0]{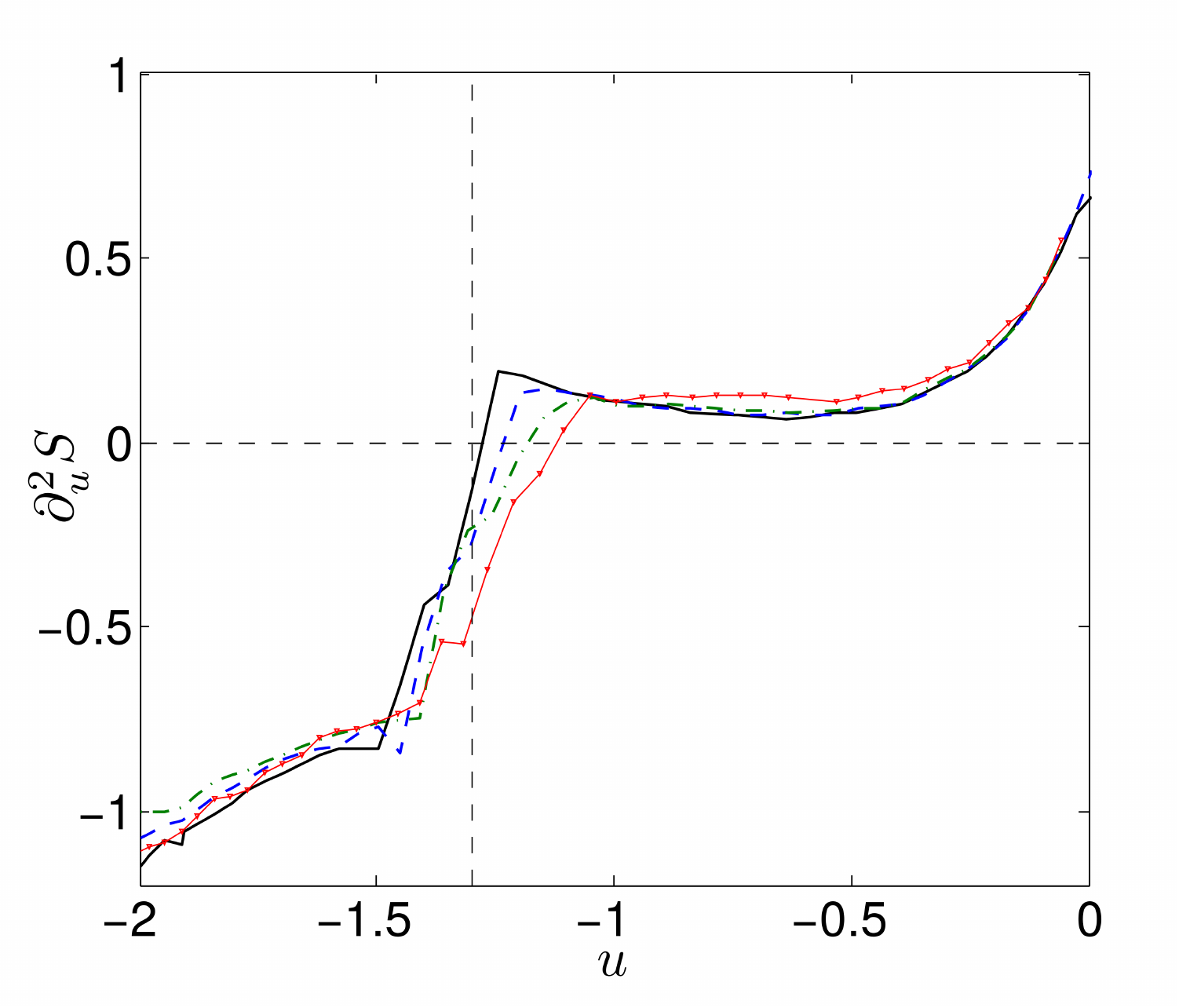}
\caption{(Color online) Second derivative $\partial^2S/\partial u^2$ of the configurational entropy versus the average potential energy per degree of freedom $u$.
Lattice dimensions: $n^3=6\times 6\times 6$ (thin solid line with small triangles), $n^3=8\times 8\times 8$  (dot-dashed line),   $n^3=10\times 10\times 10$ (dashed line), $n^3=14\times 14\times 14$ (thick solid line). }
\label{deriv2S}
\end{figure}

From the numerical results concerning the functions $T(E)$ and $u(E)$, reported in \autoref{1} and \autoref{intern-energ}, respectively, we  computed the first and second derivatives of the configurational entropy as
\begin{equation}\label{derivprima}
\frac{\partial S}{\partial u} = \frac{\partial S}{\partial E}\frac{dE}{du} = \frac{1}{T(E)}\frac{dE}{du}\ ,
\end{equation}
\begin{equation}\label{derivsecond}
\frac{\partial^2S}{\partial u^2} = \frac{\partial}{\partial u} \left(\frac{1}{T(E)}\frac{dE}{du}\right)\ .
\end{equation}
The derivative $(dE/du)$ entering Eq.\eqref{derivprima} is obtained after inversion of the function $u=u(E)$ reported in \autoref{intern-energ} and by means of a spline interpolation of its points. 
Whereas $\partial^2_uS(u)$ in Eq.\eqref{derivsecond} is computed from the raw numerical data, and the derivatives with respect to $u$ have been obtained by means of a standard central difference formula.

The four patterns of $\partial_uS(u)$, computed for different sizes of the lattice and  reported in \autoref{deriv1S} and \autoref{deriv1Szoom}, show that each $\partial_uS(u)$ appears splitted into two monotonic branches, one decreasing and the other increasing as functions of $u$, respectively. Approximately out of the interval $u\in(-1.6, -0.65)$ the four patterns are perfectly superposed, whereas within this interval - which contains the transition value $u_c\simeq -1.32$ - we can observe that the transition from $\partial_uS<0$ to $\partial_uS>0$ gets sharper at increasing lattice dimension. This means that the second derivative of the entropy, $\partial^2_uS(u)$, tends to make a sharper jump at increasing $N$. And in fact, this is what is suggested by the four patterns of $\partial^2_uS(u)$ - computed for the same sizes of the lattice -  reported in \autoref{deriv2S}. These are strongly suggestive to belong to a sequence of patterns converging to a step-like limit pattern. In this case the third order derivative $(\partial^3S/\partial u^3)$ would asymptotically diverge entailing a loss of analyticity of the entropy which, in fact, would drop to  $S_\infty(u)\in {\cal C}^1$.
And this is in agreement with the above proposed classification  {\it \`a} {\it la} Ehrenfest. 

It is worth noting that we have found evidence of only one transition point, despite the presence of two peaks of the specific heat, thus confirming that in order to correctly characterize  a phase transition in the microcanonical ensemble one has to look for the signals of analyticity loss of the entropy.

\subsection{A geometric observable for the level sets $\Sigma_v$ in configuration space}
We have seen in the preceding Section that - within the confidence limits of a numerical investigation - the first order phase transition of the gauge model under investigation seems to correspond to an asymptotic divergence of the third derivative $(\partial^3S/\partial u^3)$ of the microcanonical configurational entropy. Under the main theorem in Ref.\cite{NPB2} this should stem from a topological change of the potential level sets $\Sigma_u = V^{-1}(u)$. In order to get some information of topological kind about these level sets one has to resort to concepts and methods of differential topology. In fact, differential topology allows to catch some topological information on differentiable manifolds through suitable curvature integrals (the Gauss-Bonnet theorem being the first classical example of this type).
Relevant geometric quantities can be computed through the extrinsic geometry of hypersurfaces of a Euclidean space. To do this one has to study the way in which an $N$-surface 
$\Sigma$ curves around in ${\mathbb R}^{N+1}$ by measuring the way the normal direction
changes as we move from point to point on the surface. The rate of change
of the normal direction ${\bf N}$ at a point $x\in\Sigma$ in
direction ${\bf v}$ is described by the {\it shape operator} (sometimes also
called Weingarten's map) $L_x({\bf v}) = - \nabla_{\bf v}{\bf N}= -({\bf v}\cdot\nabla){\bf N}$, where
${\bf v}$ is a tangent vector at $x$ and $\nabla_{\bf v}$ is the directional
derivative; gradients and vectors being represented in ${\mathbb R}^{N+1}$.

For the level sets of a regular function, as is the case of the constant-energy
hypersurfaces in the phase space of Hamiltonian systems or of the equipotential
hypersurfaces in configuration space, thus generically defined through a
regular real-valued function $f$ as $\Sigma_a:=f^{-1}(a)$, the normal vector
is ${\bf N}= \nabla f/\Vert\nabla f\Vert$.
The eigenvalues $\kappa_1(x),\dots,\kappa_N(x)$ of the shape operator are the principal curvatures at $x\in\Sigma$. 
For the potential level sets $\Sigma_v = V^{-1}(v)$ the trace of the shape operator at any given point is the mean curvature at that point and can be written as \cite{book,thorpe}
\begin{equation}\label{emme}
M = - \frac{1}{N}\nabla\cdot\left(\dfrac{\nabla V}{\Vert\nabla V\Vert}\right)=  \frac{1}{N}\sum_{i=1}^N\kappa_i\ .
\end{equation}
We have numerically computed the second moment of $M$  averaged along the Hamiltonian flow 
\begin{equation}\label{sigmaM}
\sigma_M = N \langle Var (M)\rangle_t =  N [\langle M^2\rangle_t - \langle M\rangle_t^2] \simeq \frac{1}{N} \sum_{i=1}^N\langle \kappa_i^2\rangle_t  - \frac{1}{N} \sum_{i=1}^N\langle \kappa_i\rangle_t ^2\ ,
\end{equation}
where we have assumed that the correlation term $N^{-2}\sum_{i,j}[\langle k_ik_j\rangle_t - \langle k_i\rangle_t\langle k_j\rangle_t]$ vanishes. In fact, on the one side there is no conserved ordering of the eigenvalues of the shape operator along a dynamical trajectory, and on the other side the averages are performed along chaotic trajectories (the largest Lyapounov exponent is always positive) so that $k_i$ and $k_j$ vary almost randomly from point to point and independently one from the other. 
\begin{figure}[h!]
\includegraphics[width=10cm,angle=0]{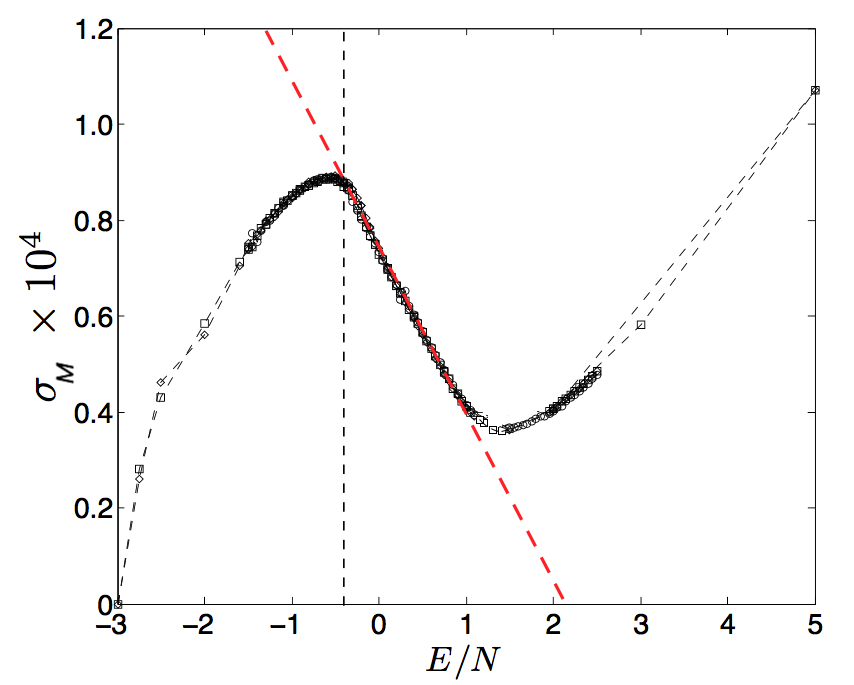}
\caption{(Color online) Second moment of the total mean curvature of the potential level sets $\Sigma_u$ versus energy density $E/N$. 
$n^3=6\times 6\times 6$ (rhombs),   $n^3=8\times 8\times 8$  (squares), $n^3=10\times 10\times 10$ (circles). The oblique dashed line is a guide to the eye. The vertical dashed line corresponds to the point where the second derivative  $d^2\sigma_M/dE^2$ jumps from a negative value to zero.}
\label{fluttM-E}
\end{figure}

\begin{figure}[h!]
\includegraphics[width=9cm,angle=0]{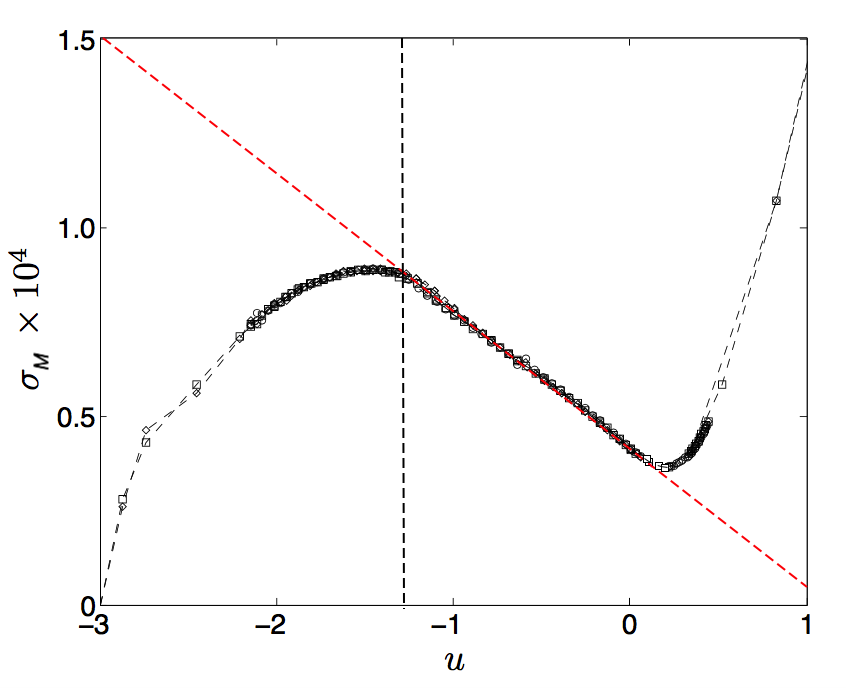}
\caption{(Color online) Second moment of the total mean curvature of the potential level sets $\Sigma_u$ versus the average potential energy per degree of freedom $u$. 
$n^3=6\times 6\times 6$ (rhombs),   $n^3=8\times 8\times 8$  (squares), $n^3=10\times 10\times 10$ (circles). The oblique dashed line is a guide to the eye. The vertical dashed line corresponds to the point where the second derivative  $d^2\sigma_M/du^2$ jumps from a negative value to zero.}
\label{fluttM-u}
\end{figure}

The numerical results are reported in \autoref{fluttM-E} and \autoref{fluttM-u}, where an intriguing feature of the patterns of $\sigma_M (E)$ and $\sigma_M (u)$ is evident: below the transition point (marked with a vertical dashed line) the concavity of both $\sigma_M (E)$ and $\sigma _(u)$ is oriented downward so that  $d^2\sigma_M/dE^2$ and $d^2\sigma_M/du^2$ are negative,  whereas just above the transition point
both $\sigma_M (E)$ and $\sigma_M (u)$ are segments of a straight line, so that $d^2\sigma_M/dE^2$ and $d^2\sigma_M/du^2$ vanish.
Thus both derivatives make a jump at the transition point. Again within the validity limits of numerical investigations, this means that the third order derivatives, and in particular $d^3\sigma_M/du^3$, diverge. It is then natural to think of a connection with the asymptotic divergence of $d^3S/du^3$ suggested  by the results reported in the preceding Section.

\color{black}
\noindent{\underline {\textit Remark.}} There is a point of utmost importance to comment on. In presence of a phase transition (and of a finite size of a phase transition as is the case of numerical simulations) the typical variations of many observables at the  transition point are the effects of the singular properties of the statistical measures and hence of the corresponding thermodynamic potentials (entropy, free energy, pressure). But this is not true for the geometric quantity $\sigma_M (u)$ which is {\it independent} of the properties of any statistical measure. Peculiar changes of the geometry - and possibly of the topology - of the potential level sets of configuration space (detected by $\sigma_M$) constitute the deep origin, the {\it cause} of phase transitions, {\it not an effect}.  The singular pattern of  $\sigma_M$ at the transition point is a primitive phenomenon. In other words, geometrical/topological variations of the spaces where the statistical measures are defined (phase space and configuration space) entail their singular properties \cite{book}. The vice versa is meaningless.
\color{black}

Let us now see how the jump of the second derivative of $\sigma_M (u)$ and the jump of the second derivative of the configurational entropy $S(u)$ can be both attributed to a deeper phenomenon: a suitable change of the topology of the $\{\Sigma_u\}_{u\in{\mathbb R}}$. In what follows we resort to the best, non-trivial approximations at present available. 

Consider the pointwise dispersion of the principal curvatures
\begin{equation}\label{sigmaK}
s_\kappa = \frac{1}{N}  \sum_{i=1}^N(\kappa_i - {\overline{\kappa}})^2
\end{equation}
where 
\begin{equation}\label{meancurv}
{\overline{\kappa}} = \frac{1}{N}  \sum_{i=1}^N \kappa_i
\end{equation}
equation \eqref{sigmaK} is equivalently rewritten as \cite{deform}
\begin{equation}\label{kikj}
s_\kappa = \frac{1}{N^2}  \sum_{i,j=1}^N(\kappa_i - \kappa_j)^2
\end{equation}
and the time average along the Hamiltonian flow of $s_\kappa$ is then equivalently written as
\begin{equation}\label{kikj}
\langle s_\kappa\rangle_t = \frac{1}{N} \sum_{i=1}^N\langle \kappa_i^2\rangle_t   - \langle{\overline{\kappa}}^2\rangle_t  = \frac{1}{N^2}  \sum_{i,j=1}^N\langle(\kappa_i - \kappa_j)^2\rangle_t \ .
\end{equation}
Now, from Eqs.\eqref{sigmaM} and \eqref{kikj} we get 
\begin{equation}\label{differ}
\sigma_M - \langle s_\kappa\rangle_t \simeq  - \frac{1}{N} \sum_{i=1}^N\langle \kappa_i\rangle_t ^2 + \langle{\overline{\kappa}}^2\rangle_t  
\end{equation}
so that, if we make a ``mean field''-like approximation in the first term of the r.h.s. of \eqref{differ} by replacing the $\kappa_i$ with their average ${\overline{\kappa}}$ it follows 
$ \sigma_M - \langle s_\kappa\rangle_t \simeq - \langle{\overline{\kappa}}\rangle_t^2 +\langle{\overline{\kappa}}^2\rangle_t$, and as 
$\overline{\kappa}$ in Eq.\eqref{meancurv} is the same of $M$ in Eq.\eqref{emme}  one trivially gets
\begin{equation}\label{differzero}
(1 - \frac{1}{N}) \sigma_M - \langle s_\kappa\rangle_t \simeq 0
\end{equation}
so that, under this ``mean field''-like approximation, $\sigma_M$ in Eq.\eqref{sigmaM} can be used to estimate $\langle s_\kappa\rangle_t$.
 Then, as the ergodic invariant measure of chaotic Hamiltonian dynamics is the microcanonical one, the time averages $\langle\cdot\rangle_t$ provide the values of the surface averages  $\langle\cdot\rangle_{\Sigma_E}$. Hence, and within the validity limits of the undertaken approximations, an interesting connection can be used between extrinsic curvature properties of an hypersurface of a Euclidean space 
${\mathbb R}^{N+1}$ and its Betti numbers (the diffeomorphism-invariant dimensions of the cohomology groups of the hypersurface, thus topological invariants) \cite{nakahara}. This connection is  made by Pinkall's inequality given in the following.

Denoting by
\[
\sigma (L_x)^2 = \frac{1}{N^2}\sum_{i<j} (\kappa_i - \kappa_j)^2
\]
the dispersion of the principal curvatures of the hypersurface, then after Pinkall's theorem  \cite{pinkall}
\beq
\frac{1}{{\rm vol}({\mathbb S}^N)}\int_{\Sigma_v^N}[\sigma (L_x)]^N\
d\mu(x) \geq \sum_{i=1}^{N-1}\left( \frac{i}{N-i}\right)^{N/2-i}\ b_i(\Sigma_v^N)\ ,
\label{Pinkall}
\eeq
where  $b_i(\Sigma_v^N)$ are the Betti numbers of the manifold $\Sigma_v^N$ immersed in the Euclidean space ${\mathbb R}^{N+1}$,  ${\mathbb S}^N$ is an $N$-dimensional sphere of unit radius, and $\mu(x)$ is the measure on $\Sigma_v^N$.

With the help of the H\"older inequality for integrals we have
\begin{equation}\label{curv-top1}
\int_{\Sigma_v^N}[\sigma (L_x)]^2\
d\mu(x) \le \left[ \int_{\Sigma_v^N} \{[\sigma (L_x)]^2\}^{N/2} d\mu(x)\right]^{2/N}\left[ \int_{\Sigma_v^N} d\mu(x)\right]^{1/(1-2/N)}
\end{equation}
whence, at large $N$,
\begin{equation}\label{curv-topA}
\left[ \int_{\Sigma_v^N} d\mu(x)\right]^{-1} \int_{\Sigma_v^N}[\sigma (L_x)]^2\
d\mu(x) \le \left[ \int_{\Sigma_v^N} \{[\sigma (L_x)]^2\}^{N/2} d\mu(x) \right]^{2/N} 
\end{equation}
this inequality becomes an equality when $\vert f\vert^p/\Vert f\Vert^p_p = \vert g\vert^q/\Vert g\Vert^q_q$ almost everywhere \cite{holderequal}, where $\Vert f\Vert_p$ is the standard $L^p$ norm
$\Vert f\Vert_p = \left(\int_S \vert f\vert^p d\mu(x)\right)^{1/p}$, where $S$ is a measurable space. In the inequalities above $g(x)=1$, thus the H\"older inequality becomes an equality when 
$\vert f\vert^p=\Vert f\Vert^p_p /\int_S d\mu(x)$, that is, when $\vert \sigma (L_x)\vert^N$ equals its average value almost everywhere on $\Sigma_v^N$.
Introducing a positive remainder function $r(v)$,  Eq.\eqref{curv-topA} is rewritten as
\begin{equation}\label{curv-top2}
\left[ \int_{\Sigma_v^N} d\mu(x)\right]^{-1} \int_{\Sigma_v^N}[\sigma (L_x)]^2\
d\mu(x)  = \left[ \int_{\Sigma_v^N} \{[\sigma (L_x)]^2\}^{N/2} d\mu(x) \right]^{2/N} - r(v)\ .
\end{equation}
For the model under investigation, the pointwise dispersion of the principal curvatures of the potential level sets actually displays a limited variability from point to point. This follows from the observation that the numerically computed variance of the mean curvature in Eq.\eqref{sigmaM} is very fastly convergent to its asymptotic value, independently of the initial condition \cite{notaM}. Hence, in the present case, the remainder $r(v)$ appears to be a small correction and, consequently, the H\"older inequality is tight.

Then, using 
\begin{equation}\label{curv-top3}
\sigma_M = N [\langle M^2\rangle_{\Sigma_v} - \langle M\rangle_{\Sigma_v}^2] \sim \left[ \int_{\Sigma_v^N} d\mu(x)\right]^{-1} \int_{\Sigma_v}[\sigma (L_x)]^2\
d\mu(x)  
\end{equation}
together with Eq.\eqref{curv-top2}  and the Pinkall inequality, one finally gets
\begin{equation}\label{curv-top4}
\sigma_M  \sim \left[ \int_{\Sigma_v^N} \{[\sigma (L_x)]^2\}^{N/2} d\mu(x) \right]^{2/N} - r(v) \sim \left[ {{\rm vol}({\mathbb S}^N)} \sum_{i=1}^{N-1}\left( \frac{i}{N-i}\right)^{N/2-i}b_i(\Sigma_v^N)\right]^{2/N} - r(v) \ ,
\end{equation}
that is, the observable $\sigma_M(v)$ is explicitly related with the topology of the level sets $\Sigma_v^N$.
 This relation, even being an approximate one, is definitely non-trivial because there are very few possibilities of relating total curvature properties of a manifold with its topological invariants. On the other hand, both Pinkall's inequality and the H\"older inequality are sufficiently tight to make Eq.\eqref{curv-top4} meaningful. In fact, in addition to the already given arguments concerning the H\"older inequality in Eq.\eqref{curv-topA},
Pinkall's inequality stems from the Morse inequalities $\mu_k(M)\ge b_k(M)$ which relate the Morse indexes $\mu_k(M)$ to the Betti numbers $b_k(M)$ of a manifold $M$ (Pinkall's inequality would be replaced by an equality if written with Morse indexes), and these Morse inequalities are very tight since the alternating sums of Morse indexes and of Betti numbers, respectively, give the same result (the Euler characteristic). Therefore, 
the integral in the l.h.s. of Eq.\eqref{curv-top4} necessarily follows the topological variations of  the $\Sigma_v^N$ described by the weighted sum of its Betti numbers. 
The consequence is that a suitable variation with $v$ of the weighted sum of the Betti numbers of a $\Sigma_v^N$ can be sufficient to entail a sudden change of the convexity of the function $\sigma_M(v)$, as reported in \autoref{fluttM-u}, and thus entail a discontinuity of its second derivative \cite{nota}.

On the other hand, the existence of a relationship between thermodynamics and configuration space topology is provided by the following exact formula  \cite{book,NPB2}
\[
S_N^{(-)}(v) =({k_B}/{N}) \log \left[ \int_{M^N_v}\ d^Nq\right] 
\]
\begin{equation}
=\frac{k_B}{N} \log \left[ vol
[{M^N_v \setminus\bigcup_{i=1}^{{\cal N}(v)} \Gamma(x^{(i)}_c)}]\ +
\sum_{i=0}^N w_i\ \mu_i (M^N_v)+ {\cal R} \right]  ,\label{exactS}
\end{equation}
where $S^{(-)}_N$ is the configurational entropy,  and the
$\mu_i(M^N_v)$ are the Morse indexes (in one-to-one correspondence
with topology changes) of the submanifolds $\{ M^N_{v}=V_N^{-1}((-\infty,v])\}_{v \in{\mathbb R}}$ 
of configuration space; in square brackets: the first term is the result of the excision of certain neighbourhoods 
of the critical points of the interaction potential from  $M^N_{v}$; the second term
is a weighed sum of the Morse indexes, and the third term is a smooth function of $N$ and $v$.
Again, sharp changes in the potential energy pattern of at least some of
the $\mu_i(M^N_v )$ (thus of the way topology changes with $v$) affect $S_N^{(-)}(v)$ and its
derivatives. 

In other words, both the jump of the second derivative of the entropy and of the second derivative of 
$\sigma_M$ are possibly rooted in the same topological ground, where some adequate variation of the topology of the $\Sigma_v^N$ - foliating the configuration space - takes place.
Notice that even if in Eq.\eqref{exactS} $S_N^{(-)}(v)$ depends on the topology of the $M_v^N$ through the Morse indexes $\mu_i (M^N_v)$, in the framework of Morse theory a topology change of a level set $\Sigma_v^N$ is always associated with a topology change of the associated manifold $M_v^N$ of which 
$\Sigma_v^N$ is the boundary \cite{milnor}. 

Summarizing,   the topology changes indirectly detected by the function $\sigma_M(u)$ can affect the configurational entropy $S_N(v)$ and its tendency to develop an asymptotic  discontinuity 
of $\partial_v^2S_\infty(v)$ (we use $u$ and $v$ interchangeably).

Finally, in Appendix we show that the non-trivial contribution to the homology groups of the energy level sets $\Sigma_E^N$ comes from the homology groups of the configuration space submanifolds $M_v^N\subset M_E^N$ and $\Sigma_v^N\subset \Sigma_E^N$.  Therefore, the topology variations of the $\Sigma_v^N$ imply topology variations of the $\Sigma_E^N$, and these  necessarily affect also the functional dependence on $E$ of the total entropy $S_N(E)$. In fact, the variation with $v$ of the topology of the $\Sigma_v^N$ is in one-to-one correspondence with some variation with $v$ of the Betti numbers $b_i(\Sigma_v^N)$ entering Eq.\eqref{curv-top4}, and this entails the variation with $E$ of the Betti numbers $b_i(\Sigma_E^N)$, so that, according to the following formula \cite{book} for the total entropy  
\begin{equation}
S_N(E)\approx  \frac{k_{\rm B}}{N} \log \left[
{\rm vol}({\mathbb S}_1^{N-1})\sum_{i=0}^N b_i(\Sigma_E^N)
+ {\cal R}_1(E)\right] + {\cal R}_2(E) \ ,
\label{entropia}
\end{equation}
where ${\cal R}_1(E)$, and ${\cal R}_2(E)$ are smooth functions, we see that the variation with $v$ of the topology of the $\Sigma_v^N$ implies also the variation with $E$ of the total entropy.

\section{Concluding remarks}
We have tackled the problem of characterizing a phase transition in the absence of global symmetry breaking from the point of view of Hamiltonian dynamics and related geometrical and topological aspects. In this condition the Landau classification of phase transitions does not apply, because no order parameter - commonly associated with a global symmetry - exists. The system chosen is inspired by the dual of the Ising model, and the discrete variables are replaced by continuous ones. We stress that our work has nothing to do with the true dual Ising model, which has just suggested how to define a classical Hamiltonian system with a local (gauge) symmetry. 
Since the ergodic invariant measure for generically non-integrable Hamiltonian systems is the microcanonical measure in phase space, studying phase transitions through Hamiltonian dynamics is the same as studying them in the microcanonical ensemble. 

A standard analysis has been performed to locate the phase transition and to determine its order through the shape of the caloric curve, $T=T(E)$, which appeared  typical of a first order phase transition. The presence of energy intervals of negative specific heat are indicative of ensemble nonequivalence. At variance with what has been systematically observed for systems undergoing symmetry-breaking phase transitions, the energy pattern of the largest Lyapunov exponent does not allow to locate the transition point. 

After the Yang-Lee theory, phase transitions are commonly associated with a loss of analyticity of a thermodynamic potential entailing non-analytic patterns of thermodynamic observables (the pertinent potential depends on the statistical ensemble chosen). However, the caloric curve found for our gauge model is very regular, no bifurcating order parameter exists, and the peaks of specific heat are just due to horizontal tangencies to the caloric curve. In other words, apparently there is no evidence of the existence of  genuinely non-analytic energy pattern of some observable. However, by directly looking at the energy pattern of the entropy, we have identified a point of discontinuity of its second derivative (or at least a finite size version of such a discontinuity), hence an asymptotic divergence of the third derivative of the entropy. Then we have discussed how this fits into a classification scheme {\it \`a} {\it la} Ehrenfest, adapted to the framework of microcanonical ensemble, that allows to determine the order of a phase transition on the basis of the differentiability class of the entropy in the thermodynamic limit.

Remarkably, we have found a quantity,  $\sigma_M(v)$, that - by measuring the total degree of inhomogeneity of the extrinsic curvature of the potential level sets $\Sigma_v=V^{-1}(v)$ in configuration space - identifies the phase transition point.  This quantity is not a thermodynamic observable, has a purely geometric meaning, and displays a discontinuity of its second derivative in coincidence with the same kind of discontinuity displayed by the entropy. Rather than being a trivial consequence of the presence of the phase transition, the peculiar change of the geometry of the $\{\Sigma_v^N\}_{v\in{\mathbb R}}$ so detected is the deep cause of the singularity of the entropy. In fact,  the potential level sets are simply subsets of ${\mathbb R}^N$ defined as  $\Sigma_v^N =\{(q_1,\dots,q_N)\in {\mathbb R}^N\vert V(q_1,\dots,q_N) = v\}$, whose ensemble $\{\Sigma_v^N\}_{v\in{\mathbb R}}$ foliates the configuration space; the volume $\Omega(v,N)$ - of each leaf $\Sigma_v^N$ - and the way it varies as a function of $v$ is just a matter of geometrical/topological properties of the leaves of the foliation. These properties entail the $v$-dependence of the entropy $S_N(v) =(1/N) \log \Omega(v,N)$, and, of course, its differentiability class. This is why the $v$-pattern of the quantity $\sigma_M(v)$ is not the consequence of the presence of a phase transition but, rather,  the reason of its appearance. This is already a highly non trivial fact indicating that whether a physical system can undergo a phase transition is somehow already encoded in the interactions among its degrees of freedom described by the potential function $V(q_1,\dots,q_N)$, independently of the statistical ensemble chosen to describe its macroscopic observables. However, we can wonder if one can go deeper by looking for the origin of the peculiar changes with $v$ of the geometry of the $\Sigma_v^N$. Actually,  by resorting to a theorem in differential topology, and with some approximations, these geometrical changes  appear to be due to  changes of the topology of both the potential level sets in configuration space and  the energy level sets in phase space.
Therefore, the results of the present work lend further support to the topological theory of phase transitions. 

Moreover, since the practical computation of $\sigma_M(v)$, or of $\sigma_M(E)$, is rather straightforward, this can be used to complement the study of transitional phenomena in the absence of symmetry-breaking, as is the case of: liquid-gas change of state, Kosterlitz-Thouless transitions, glasses and supercooled liquids,  amorphous and disordered systems, folding transitions in homopolymers and proteins, both classical and quantum transitions in small $N$ systems. 
{\color{black}
With respect to the latter case, a remark about the topological theory is in order. In nature, phase transitions (that is major qualitative physical changes) occur also in very small systems with $N$ much smaller than the Avogadro number,  but their mathematical description through the loss of analyticity of thermodynamic observables requires the asymptotic limit $N\to\infty$. To the contrary, within the topological framework a sharp difference between the presence or the absence of a phase transition can be made also at any finite and even very small $N$. At finite $N$, the microscopic states that significantly contribute to the statistical averages of thermodynamic observables are spread in regions of configuration space which get narrower as $N$ increases, so that the statistical measures better concentrate on a specific potential level set thus better detecting its sudden and major topology changes, if any. Eventually, in the $N\to\infty$ limit the extreme sharpening of the effective support of the measure leads to a topology-induced nonanalyticity of thermodynamic observables \cite{book}.}

Furthermore, even if somewhat abstract, the model studied in the present work has the basic properties of a lattice gauge model, that is, its potential depends on the circulations of the gauge field on the plaquettes, so that the geometrical/topological approach developed here could be also of some 
interest to the numerical investigation of phase transitions of Euclidean gauge theories on lattice. In fact, computing $\sigma_M(v)$, or $\sigma_M(E)$, is definitely easier than computing the Wilson loop, commonly adopted in place of an order parameter for gauge theories.
Actually, a few decades ago, several papers on the microcanonical formulation of quantum field theories appeared \cite{mQFT0,mQFT5}, motivated by the fact that in statistical mechanics and in field theory there are systems for which the canonical description is pathological, 
but the microcanonical is not, also arguing, for instance and among other things, that a microcanonical formulation of quantum gravity may  be less pathological than the usual canonical formulation
\cite{mQFT1,mQFT2,mQFT3,mQFT4}. More recent works can also be found on these topics \cite{mQFT6,mQFT7,mQFT8,mQFT9}. 

Finally, as a side issue, it is provided here an example of statistical ensemble non-equivalence in a system with short-range interactions. Ensemble non-equivalence is another topic which is being given much attention in recent literature \cite{campa}.

\vfil

\section*{Appendix}
\subsection{Relation between topological changes of the $\Sigma_v$ and of the $\Sigma_E$}
Now, let us see why a topological change of the configuration space submanifolds $\Sigma_v=V^{-1}(v)$ (potential level sets)  implies the same phenomenon for the $\Sigma_E$. The potential level sets are the basic objects, foliating configuration space, that  represent the nontrivial topological part of phase space. The link of these geometric objects with microcanonical entropy is given by 
\begin{equation}
S^{(-)}(E) =\frac{k_B}{2N} \log \int_0^E d\eta \int d^Np\  \delta (\sum_{\bf i} p_{\bf i}^2/2 - \eta ) \int_{\Sigma_{E-\eta}} \frac{d\sigma}{\Vert\nabla V\Vert} \ .
\end{equation}
As $N$ increases the microscopic configurations giving a relevant contribution to the entropy, and to any  microcanonical average, concentrate closer and closer on the level set $\Sigma_{\langle E-\eta\rangle}$.
A link among the topology of the energy level sets and the topology of configuration
space can be established for systems described by a Hamiltonian of the form $\mathcal{H}_N(p,q)=\sum_{i=1}^N p_i^2/2+V_N(q_1,...,q_N)$. 

In fact, (using a cumbersome notation for the sake of clarity) the level sets $\LevelSetfunc{E}{{\mathcal{H}}_N}$ of the energy function ${\mathcal{H}}_N$  can be given by the disjoint union of a trivial unitary sphere bundle (representing the phase space region where the kinetic energy does not vanish) and the hypersurface in configuration space where the potential energy takes the total energy value (details are given in \cite{refine})
\begin{equation}
\LevelSetfunc{E}{{\mathcal{H}}_N} {\mathrm{\ homeomorphic\  to}\ } \Mfunc{E}{{V}_N}\times \mathbb{S}^{N-1}\,\,\,\bigsqcup\,\,\,\LevelSetfunc{{E}}{{V}_N}
\label{eq:Structure_sigmaE}
\end{equation}
where $\mathbb{S}^n$ is the $n$-dimensional unitary sphere and 
\begin{equation}
\begin{split}
&\Mfunc{c}{f}=\left\{x\in\mathrm{Dom}(f)|f(x)< c\right\}, \\ \\
 &\LevelSetfunc{c}{f}=\left\{x\in\mathrm{Dom}(f)|f(x)=c\right\}.
 \end{split}
\end{equation}
The idea that finite $N$ topology, and "asymptotic topology" as well, of $\LevelSetfunc{{E}}{{\mathcal{H}}_N}$ is affected by the  topology of the accessible region of configuration space is suggested by the \textit{K\"unneth formula}: if $H_{k}(X)$ is the $k$-th homological group of the topological space $X$ on the  field $\mathbb{F}$ then
\begin{equation}
H_{k}(X\times Y;\mathbb{F})\simeq\bigoplus_{i+j=k} H_{i}(X;\mathbb{F})\,\,\otimes\,\, H_{j}(Y;\mathbb{F})\,\,.
\end{equation}
Moreover, as $H_{k}\left(\sqcup_{i=1}^N X_{i},\mathbb{F}\right)=\bigoplus_{i}^{N}H_{k}(X_{i},\mathbb{F})$, it follows that:
\begin{equation}
\begin{split}
&H_{k}\left(\LevelSetfunc{{E}}{{\mathcal{H}}_N},\mathbb{R}\right) \label{homolog}\\
&\simeq\bigoplus_{i+j=k} H_{i}\left(\Mfunc{{E}}{{V}_N};\mathbb{R}\right)\otimes\,\, H_{j}\left(\mathbb{S}^{N-1};\mathbb{R}\right)\oplus H_{k}\left(\LevelSetfunc{{E}}{{V}_N};\mathbb{R}\right)\\
&\simeq  H_{k-(N-1)}\left(\Mfunc{{E}}{{V}_N};\mathbb{R}\right)\otimes \mathbb{R}\oplus H_{k}\left(\Mfunc{{E}}{{V}_N};\mathbb{R}\right)\otimes \mathbb{R}\\
&\oplus H_{k}\left(\LevelSetfunc{{E}}{{V}_N};\mathbb{R}\right) 
\end{split}
\end{equation}
the r.h.s. of Eq.(\ref{homolog}) shows that the topological changes of $\LevelSetfunc{{E}}{{\mathcal{H}}_N}$ only stem from the topological changes in configuration space.


\begin{references}


\bibitem{marco} Part of this work was done while M.P. was on leave of absence from Osservatorio
Astrofisico di Arcetri, Florence, Italy.

\bibitem{YL} C.N. Yang, and T.D. Lee, {\it Statistical theory of equations of state and phase transitions I. Theory of
condensation}, Phys. Rev. {\bf 87}, 404 - 409 (1952); T.D. Lee, and C.N. Yang,
{\it Statistical theory of equations of state and phase transitions II. Lattice gas
and Ising model}, Phys. Rev. {\bf 87}, 410 - 419 (1952).

\bibitem{DLR} A comprehensive account of the Dobrushin-Lanford-Ruelle theory and of its developments can be found in: 
H.O. Georgii, {\it Gibbs Measures and Phase Transitions}, Second Edition, (De Gruyter, Berlin 2011).

\bibitem{gross} D.H.E. Gross, {\it Microcanonical Thermodynamics. Phase Transitions in ``Small'' Systems}, (World Scientific, Singapore, 2001).

\bibitem{bachmann} M. Bachmann, {\it Thermodynamics and Statistical Mechanics of Macromolecular Systems}, (Cambridge University Press, New York, 2014).

\bibitem{chomaz} Ph. Chomaz, V. Duflot, and F. Gulminelli, {\it Caloric Curves and Energy Fluctuations in the Microcanonical Liquid-Gas Phase Transition}, Phys. Rev. Lett. {\bf 85}, 3587 - 3590 (2000).

\bibitem{book} M. Pettini, {\it Geometry and Topology in Hamiltonian Dynamics and
Statistical Mechanics}, IAM Series n.33, (Springer, New York, 2007).

\bibitem{NCRev} L. Casetti, M. Cerruti-Sola, M. Modugno, G. Pettini, M. Pettini and R. Gatto,
 {\sl Dynamical and Statistical properties of Hamiltonian systems
             with many degrees of freedom}, Rivista del Nuovo Cimento {\bf 22}, 1-74 (1999), and references quoted therein.

\bibitem{physrep} L.Casetti, M. Pettini, E.G.D. Cohen, {\sl Geometric approach to Hamiltonian dynamics and
            statistical mechanics}, Phys. Rep. {\bf 337},
                  237-342 (2000), and references quoted therein.

\bibitem{Pettini} M. Pettini, {\sl  Geometrical hints for a nonperturbative approach to Hamiltonian dynamics}, Phys. Rev. E {\bf 47}, 828 (1993).

\bibitem{exact1} L. Casetti, E.G.D. Cohen, and M. Pettini, {\sl Topological origin of the phase transition in a mean-field model}, Phys. Rev. Lett. {\bf 82}, 4160 (1999).

\bibitem{exact2} L. Casetti, E.G.D. Cohen, and M. Pettini, {\sl Exact  result on topology and phase transitions at any finite $N$}, Phys. Rev. E {\bf 65}, 036112 (2002).

\bibitem{exact3} L. Casetti, M. Pettini and E.G.D. Cohen, {\sl Phase transitions and topology changes in configuration space}, J. Stat. Phys. {\bf 111}, 1091 (2003). 
        
\bibitem{exact4}  L. Angelani, L. Casetti, M. Pettini, G. Ruocco, and F. Zamponi, {\sl Topology and Phase Transitions: from an exactly solvable model to a 
relation between topology and thermodynamics}, Phys. Rev. E{\bf 71}, 036152 (2005).

\bibitem{santos}  F. A. N. Santos, L. C. B. da Silva, and M. D. Coutinho-Filho, {\sl Topological approach to microcanonical thermodynamics and phase transition of interacting classical spins}, J. Stat. Mech.  {\bf 2017}, 013202 (2017).

\bibitem{prl1} R. Franzosi, and M. Pettini, {\sl Theorem on the origin of Phase Transitions}, Phys. Rev. Lett. {\bf 92}, 060601 (2004).Ä

\bibitem{NPB1} R. Franzosi, M. Pettini, and L. Spinelli, {\sl Topology and Phase Transitions I. Preliminary results},
               Nucl. Phys. B{\bf 782} [PM], 189 (2007).
               
\bibitem{NPB2} R. Franzosi and M. Pettini, {\sl Topology and Phase Transitions II. Theorem on a necessary relation}, Nucl. Phys. B{\bf 782} [PM], 219 (2007).

\bibitem{kastner} Even though a counterexample was given in: M. Kastner and D. Mehta, {\it Phase Transitions Detached from Stationary Points of the Energy Landscape}, Phys. Rev. Lett. \textbf{107}, 160602 (2011), the problem can be fixed with a refinement of the hypotheses of the theorems, as shown in: M. Gori, R. Franzosi, and M. Pettini, 	{\sl Toward a refining of the topological theory of phase transitions}, arXiv:1706.01430 [cond-mat.stat-mech].

\bibitem{kogut} J. Kogut, {\it An introduction to lattice gauge theory and spin systems}, Rev. Mod. Phys. {\bf 51}, 659 (1979).

\bibitem{elitzur} S. Elitzur, {\it  Impossibility of Spontaneously Breaking Local Symmetries}, Phys. Rev. D{\bf 12}, 3978 (1975).

\bibitem{lapo} L. Casetti, {\it Efficient symplectic algorithms for numerical simulations of Hamiltonian flows}, Physica Scripta {\bf 51}, 29 (1995).

\bibitem{pearson} E. M. Pearson, T. Halicioglu, and W.A. Tiller, {\it Laplace-transform technique for deriving thermodynamic equations from the classical 
microcanonical ensemble}, Phys. Rev. A{\bf 32}, 3030 (1985).

\bibitem{ergodic} For generic quasi-integrable systems, in the form $H(\alpha, J) = H_0(J) + \varepsilon H_1(\alpha, J)$ with $(\alpha, J)$ action-angle coordinates, with three or more degrees of freedom, after the Poincar\'e-Fermi theorem for any $\varepsilon>0$ all the integrals of motion except the energy are destroyed, so that there is no topological obstruction to ergodicity. On the other hand, a lack of ergodicity stemming from KAM theorem requires exceedingly tiny values of the perturbation and $\varepsilon <\varepsilon_c$ where $\varepsilon_c$ drops to zero more than exponentially with the number of degrees of freedom. Moreover, generic nonintegrable systems are chaotic, so that, from the physicists' viewpoint these systems are {\it bona fide} ergodic and mixing. 

\bibitem{cccp} L. Caiani, L. Casetti, C. Clementi, and M. Pettini, {\sl Geometry of dynamics, Lyapunov exponents and phase transitions}, 
Phys. Rev. Lett. {\bf 79}, 4361 (1997).

\bibitem{ramaswami} V. Mehra, and R. Ramaswamy, {\sl Curvature fluctuations and the Lyapunov exponent at melting}, Phys. Rev. E {\bf 56}, 2508 (1997).

\bibitem{cccppg} L. Caiani, L. Casetti, C. Clementi, G. Pettini, M. Pettini,
and R. Gatto, {\sl Geometry of dynamics and phase transitions in classical lattice
          $\varphi^4$ theories}, Phys. Rev. E {\bf 57}, 3886 (1998).

\bibitem{lando} L. Caiani, L. Casetti and M. Pettini, {\sl Hamiltonian dynamics of the two-dimensional lattice $\varphi^4$ model},
                J. Phys. A: Math. Gen. {\bf 31}, 3357 (1998).

\bibitem{firpo} M.-C. Firpo, {\sl Analytic estimation of the Lyapunov exponent in a mean-field model undergoing a phase transition},
Phys. Rev. E {\bf 57}, 6599 (1998).

\bibitem{dauxois} J. Barr\'e, and T. Dauxois, {\sl Lyapunov exponents as a dynamical indicator of a phase transition}, Europhys. Lett. {\bf 55}, 164 (2001).

\bibitem{thorpe} J.A. Thorpe, {\it Elementary Topics in Differential Geometry}, (Springer-Verlag, New York 1979).

\bibitem{deform} Y. Zhang, H. Wu, and L. Cheng, {\sl Some New Deformation Formulas about Variance and Covariance}, Proceedings of 4th International Conference on
Modelling, Identification and Control, Wuhan, China, June 24-26, (2012).

\bibitem{nakahara} M. Nakahara, {\sl Geometry, Topology and Physics}, (Adam Hilger, Bristol, 1991).

\bibitem{pinkall} U. Pinkall, {\it Inequalities of Willmore Type for Submanifolds}, Math. Zeit. {\bf 193}, 241 (1986).

\bibitem{holderequal} M. Reed, and B. Simon, {\it vol. 1: Functional Analysis, revised and enlarged edition}, (Academic Press, San Diego, 1980).

\bibitem{notaM} The spread of the values of $\sigma_M = N \langle Var (M)\rangle_{\Delta t} =  N [\langle M^2\rangle_{\Delta t}- \langle M\rangle_{\Delta t}^2] $, numerically computed along short segments of  time duration $\Delta t = 100$, out of  long phase space trajectories - of a time duration of $t = 10^6$ - typically amounts to $3 - 5\%$. As a consequence, on this slightly coarse-grained manifold the condition to make H\"older inequality an equality is close to be satisfied, what indicates that in the case under study the H\"older inequality is tight.

\bibitem{nota} The Betti numbers - as well as Morse indexes - are integers so that their sum, weighted or not, forms only staircase-like patterns which do not qualify as  continuous and possibly differentiable functions. Actually the technical details of the reason why the corners of these staircase-like patterns are rounded can be found in Section 9.5 of Ref.\protect{\cite{book}}. 

\bibitem{milnor} J. Milnor, {\it Morse Theory, Ann. Math. Studies {\bf 51}}, (Princeton University Press, Princeton 1963).

\bibitem{mQFT0} D.J.E. Callaway, and A. Rahman, {\it Lattice gauge theory in the microcanonical ensemble},  Phys. Rev. D{\bf 28}, 1506 (1983).

\bibitem{mQFT5} M. Fukugita,T. Kaneko, and A. Ukawa, {\it Testing microcanonical simulation with SU(2) lattice gauge theory}, Nucl. Phys. B{\bf 270} 365 (1986).  

\bibitem{mQFT1} A. Strominger, {\it Microcanonical Quantum Field Theory}, Ann. Phys. NY {\bf 146},  419 (1983).

\bibitem{mQFT2} A. Iwazaki, {\it Microcanonical formulation of Quantum field theories},  Phys. Lett. B{\bf 141}, 342 (1984).

\bibitem{mQFT3} Y. Morikawa, and A. Iwazaki, {\it Supercooled states and order of phase transitions in microcanonical simulations}, Phys. Lett. B{\bf 165}, 361 (1984).

\bibitem{mQFT4} S. Duane, {\it Stochastic quantization versus the microcanonical ensemble: getting the best of both worlds}, Nucl. Phys. B{\bf 257} , 652 (1985).

\bibitem{mQFT6}  D.J. Cirilo-Lombardo, {\it Quantum field propagator for extended-objects in the microcanonical ensemble and the S-matrix formulation}, Phys. Lett. B{\bf 637}, 133 (2006).

\bibitem{mQFT7} R. Casadio, and B. Harms, {\it Microcanonical Description of (Micro) Black Holes}, Entropy {\bf 13}, 502 (2011).

\bibitem{mQFT8} A. Sinatra, and Y. Castin, {\it Genuine phase diffusion of a Bose-Einstein condensate in the microcanonical
ensemble: A classical field study}, Phys. Rev. A{\bf 78}, 05361 (2008).

\bibitem{mQFT9} Y. Strauss, L. P. Horwitz, J. Levitan, and A. Yahalom, {\it  Quantum field theory of classically unstable Hamiltonian dynamics}, J. Math. Phys. {\bf 56}, 072701 (2015).

\bibitem{campa} A. Campa, T. Dauxois, and S. Ruffo, {\it Statistical mechanics and dynamics of solvable models with long-range interactions}, Phys. Rep. {\bf 480}, 57 - 159 (2009).











\end{references}
\end{document}